\documentclass[fleqn,usenatbib]{mnras}

\usepackage[T1]{fontenc}
\usepackage{ae,aecompl}

\usepackage{graphicx}	% Including figure files
\usepackage{amsmath}	% Advanced maths commands
\usepackage{amssymb}	% Extra maths symbols

\usepackage[landscape]{pdflscape}

\title[MOJAVE -- XIV. Shapes and opening angles of AGN jets]{MOJAVE -- XIV. Shapes and opening angles of AGN jets}

\author[A. B. Pushkarev et al.]{
A. B. Pushkarev,$^{1,2}$\thanks{E-mail: pushkarev.alexander@gmail.com (ABP)}
Y. Y. Kovalev,$^{2,3}$
M. L. Lister,$^{4}$
T. Savolainen$^{5,6,3}$\\
$^{1}$Crimean Astrophysical Observatory, Nauchny 298409, Crimea, Russia\\
$^{2}$Astro Space Center of Lebedev Physical Institute, Profsoyuznaya 84/32, Moscow 117997, Russia\\
$^{3}$Max-Planck-Institut f\"ur Radioastronomie, Auf dem H\"ugel 69, 53121 Bonn, Germany\\
$^{4}$Department of Physics, Purdue University, 525 Northwestern Avenue, West Lafayette, IN 47907, USA\\
$^{5}$Aalto University Mets\"ahovi Radio Observatory, Mets\"ahovintie 114, 02540 Kylm\"al\"a, Finland\\
$^{6}$Aalto University Department of Electronics and Nanoengineering, PL 15500, 00076, Aalto, Finland
}

\date{Accepted 2017 April 4. Received 2017 March 12; in original form 2017 January 9}
\pubyear{2017}

\begin{document}
\label{firstpage}
\pagerange{\pageref{firstpage}--\pageref{lastpage}}
\maketitle

\begin{abstract}
We present 15 GHz stacked VLBA images of 373 jets associated with active galactic nuclei (AGN) 
having at least five observing epochs within a 20~yr time interval 1994--2015 from the MOJAVE 
programme and/or its precursor, the 2~cm VLBA Survey. These data are supplemented by 1.4~GHz 
single-epoch VLBA observations of 135 MOJAVE AGNs to probe larger scale jet structures. 
The typical jet geometry is found to be close to conical on scales from hundreds to thousands 
of parsecs, while a number of galaxies show quasi-parabolic streamlines on smaller scales. 
A true jet geometry in a considerable fraction of AGNs appears only after stacking
epochs over several years. The jets with significant radial accelerated motion undergo more
active collimation. We have analysed total intensity jet profiles transverse to the local jet 
ridgeline and derived both apparent and intrinsic opening angles of the flows, with medians 
of $21\fdg5$ and $1\fdg3$, respectively. The \textit{Fermi} LAT-detected gamma-ray AGNs 
in our sample have, on average, wider apparent and narrower intrinsic opening angle, and smaller 
viewing angle than non LAT-detected AGNs. We have established a highly significant correlation 
between the apparent opening angle and gamma-ray luminosity, driven by Doppler beaming and 
projection effects. 

\end{abstract}


\section{Introduction}

The current understanding of the phenomenon of active galactic nuclei (AGN) suggests that 
accretion of matter onto black holes with masses up to $M_\text{bh}\sim10^{10}M\odot$ is 
converted into kinetic energy of highly collimated bipolar outflows of plasma along the 
rotation axis of the black hole or accretion disk. The jets are formed in the immediate 
vicinity of the central engine and become detectable at distances of a few tens of gravitational 
radii ($R_\text{g}=GM_\text{bh}/c^2$) at millimeter wavelengths \citep{Junor99_M87,Hada11_M87}.
At scales of $10^3-10^5\,R_\text{g}$, the outflows are accelerated to relativistic speeds, as
follows from the apparent superluminal motions \citep{MOJAVE_XIII}, high Doppler factors 
\citep{Hovatta09} and extreme brightness temperatures that may significantly exceed the limit 
of $10^{12}$~K set by inverse Compton cooling \citep{Kellermann69}. The jets can propagate
up to megaparsec scales, where they become diffuse while interacting with the intergalactic medium.
The nature of the processes and the physical conditions which govern forming, accelerating and
collimating these relativistic jets remain among the most challenging problems of contemporary 
astrophysics.

By now, jet shapes have been studied only for a very limited number of nearby sources, such as 
M87 \citep{Asada12}, Mkn~501 \citep{Giroletti08_Mkn501}, Cygnus~A \citep{Boccardi15_CygnusA}, 
NGC 6251 \citep{Tseng16}, NGC 1052 \citep{Kadler04}, 3C 66A \citep{Bottcher05}, and 3C~84 
(Savolainen, private communication) to probe the innermost jet regions. In \cite{Pushkarev09} 
we analysed the opening angles of 142 AGN jets, based on single-epoch VLBA data. These AGNs 
were drawn from the MOJAVE radio flux density-limited sample referred to as MOJAVE-1 
\citep{MOJAVE_V} and an accompanying gamma-ray selected sample \citep{Lister11}. Here we present 
opening angle analysis of a larger sample comprising 373 AGNs, together with the first systematic 
study of jet shapes on parsec scales using multi-epoch VLBA observations at 15~GHz and on larger 
scales probed by single-epoch VLBA observations at 1.4~GHz of the MOJAVE-1 sources. 

This paper is part of a series based on data from the MOJAVE (Monitoring Of Jets in Active galactic 
nuclei with VLBA Experiments) programme\footnote{\url{http://www.astro.purdue.edu/MOJAVE}} to monitor 
radio brightness and polarization variations in jets associated with active galaxies with declinations 
above $-30\degr$. The earlier papers have focused on the parsec-scale kinematics of the jets 
\citep{MOJAVE_XIII,MOJAVE_X,MOJAVE_VI}, their acceleration and collimation \citep{MOJAVE_XII,MOJAVE_VII}, 
spectral distributions \citep{MOJAVE_XI}, nuclear opacity \citep{MOJAVE_IX}, Faraday rotation 
\citep{MOJAVE_VIII}, parent luminosity function \citep{MOJAVE_IV}, relativistic beaming and the 
intrinsic properties \citep{Cohen07}, kiloparsec scale morphology \citep{MOJAVE_III}, and circular 
and linear polarization \citep{MOJAVE_II,MOJAVE_I}.

The structure of this paper is as follows: In Section~\ref{s:obs}, we describe our observational data, 
source sample, stacking procedure and statistics of the stacked images; in Section~\ref{s:results}, 
we discuss our results; and our main conclusions are summarized in Section~\ref{s:summary}. We use 
the term `core'  as the apparent origin of AGN jets which commonly appears as the brightest feature 
in very long baseline interferometry (VLBI) images of blazars \citep[e.g.,][]{Lobanov_98}. We adopt 
a cosmology with $\Omega_m=0.27$, $\Omega_\Lambda=0.73$ and $H_0=71$~km~s$^{-1}$~Mpc$^{-1}$ \citep{Komatsu09}.

\section{Observational data}
\label{s:obs}
\subsection{MOJAVE programme and 2~cm VLBA Survey data. Stacked images}

\begin{table*}
\caption{Source properties. Columns are as follows:
(1) B1950 name;
(2) J2000 name;
(3) other name;
(4) \textit{Fermi} gamma-ray association name;
(5) optical class, where B = BL Lac, Q = quasar, G = radio galaxy, N = narrow-lined Seyfert 1, and U = unidentified;
(6) MOJAVE 1.5~Jy sample membership flag;
(7) redshift;
(8) reference for redshift and/or optical classification.
This table is available in its entirety in the online journal.
A portion is shown here for guidance regarding its form and content.}
\label{t:sample_properties}
\begin{tabular}{c c l c c c l l}
\hline
\noalign{\smallskip}
    B1950   &     J2000      &  \phantom{xxxx}Alias & Gamma-ray association & Opt. &  1.5 Jy  & \phantom{xx}z   &  \phantom{xx}Reference \\
      (1)   &      (2)       &  \phantom{xxxxx}(3)  &          (4)          &  (5) &     (6)  & \phantom{x\,}(7)&  \phantom{xxxxx}(8)    \\
\hline\noalign{\smallskip}
0003$+$380  &  J0005$+$3820  &  S4 0003$+$38        &  3FGL J0006.4$+$3825  &   Q  &  --      &   0.229  &  \cite{1994AAS..103..349S}  \\
0003$-$066  &  J0006$-$0623  &  NRAO 005            &  --                   &   B  &  Y       &  0.3467  &  \cite{2005PASA...22..277J} \\
0006$+$061  &  J0009$+$0628  &  CRATES J0009$+$0628 &  3FGL J0009.1$+$0630  &   B  &  --      &  --      &  \cite{2012AA...538A..26R}  \\
0007$+$106  &  J0010$+$1058  &  III Zw 2            &  --                   &   G  &  Y       &  0.0893  &  \cite{1970ApJ...160..405S} \\
0010$+$405  &  J0013$+$4051  &  4C $+$40.01         &  --                   &   Q  &  --      &   0.256  &  \cite{1992ApJS...81....1T} \\
0011$+$189  &  J0013$+$1910  &  RGB J0013$+$191     &  2FGL J0013.8$+$1907  &   B  &  --      &   0.477  &  \cite{2013ApJ...764..135S} \\
0015$-$054  &  J0017$-$0512  &  PMN J0017$-$0512    &  3FGL J0017.6$-$0512  &   Q  &  --      &   0.226  &  \cite{2012ApJ...748...49S} \\
0016$+$731  &  J0019$+$7327  &  S5 0016$+$73        &  --                   &   Q  &  Y       &   1.781  &  \cite{1986AJ.....91..494L} \\
0019$+$058  &  J0022$+$0608  &  PKS 0019$+$058      &  3FGL J0022.5$+$0608  &   B  &  --      &  --      &  \cite{2013ApJ...764..135S} \\
0026$+$346  &  J0029$+$3456  &  B2 0026$+$34        &  --                   &   G  &  --      &   0.517  &  \cite{2002AJ....124..662Z} \\
\hline
\end{tabular}
\end{table*}

For the purposes of our study, we made use of data at 15~GHz from the MOJAVE programme, the 2~cm 
VLBA Survey \citep{2cmPaperI,2002AJ....124..662Z}, and the National Radio Astronomy Observatorty 
(NRAO) data archive for those sources 
that have at least five VLBA observing epochs between 1994 August 31 and 2015 August 20 inclusive. 
In total, this produced a pool of 370 AGNs, 8 of which were excluded due to uncertainty with regards 
to the core component location at some/all epochs, resulting in a final sample of 362 sources 
with 6461 VLBA maps at 735 individual epochs. These AGNs are members of the radio flux-density 
limited MOJAVE-1 \citep{MOJAVE_V} and MOJAVE 1.5~Jy \citep{Lister15} samples, and gamma-ray flux 
limited and low-luminosity samples discussed by \cite{MOJAVE_X}. Also included is a set of weaker 
radio AGNs associated with gamma-ray hard spectrum AGNs. The compact radio flux density of all 
sources from these samples is greater than 0.1~Jy at 15~GHz. Each of the final single-epoch images 
was constructed by applying natural weighting to the visibility function. For a more detailed 
discussion of the data reduction and imaging process schemes, see \cite{MOJAVE_V}. The sample of 
362 sources is strongly dominated by flat spectrum radio quasars (67\%) and BL Lacertae objects 
(24\%), but also contains 19 radio galaxies, 5 narrow-lined Seyfert 1 galaxies, and 5 optically 
unidentified sources. The redshifts are currently known for 331 objects (91\%) ranging from 
$z_\text{min} = 0.00436$ for the galaxy 1228+126 (M87) to $z_\text{max} = 4.715$ for the quasar 
PKS 0201+113, with a mean value of the distribution close to $z = 1$. The general characteristics 
of the sources such as object name, its alternative name, gamma-ray association name, optical class, 
membership of flux density-limited MOJAVE 1.5~Jy sample, and redshift are listed in 
Table~\ref{t:sample_properties}.

\begin{figure}
\centering
\includegraphics[angle=-90,width=\columnwidth]{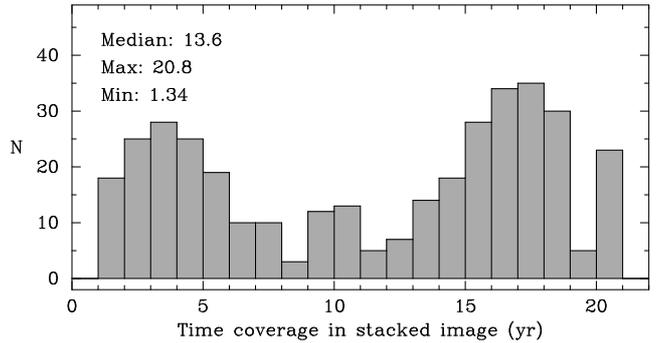}
\caption{Histograms of time interval between first and last epochs in stacked images of 362 sources.
\label{f:epochs_stat}
}
\end{figure}

Recently, \cite{MOJAVE_X} found that nearly all of the 60 most heavily observed jets over a 
decade time interval within the MOJAVE/2cm VLBA surveys displayed significant changes in 
their innermost jet position angle with time, suggesting that the superluminal AGN jet features 
seen in single-epoch images occupy only a portion of the entire jet cross-section. Thus, 
to better reconstruct the jet morphology, for each source we produced a corresponding 
stacked image using all available epochs from the MOJAVE/2cm VLBA surveys at 15~GHz convolved 
with a circular beam based on the median beam size for the source, the same pixel size 
(0.1 mas) and field of view. The stacking procedure was performed as a simple averaging in 
the image plane after aligning single-epoch total intensity maps by the VLBI core position. 
The VLBI core position was derived from the structure modeling in the $(u,v)$ domain with 
the procedure {\it modelfit} in the Caltech {\sc difmap} package \citep{difmap}. For the model 
fitting, \cite{MOJAVE_X,MOJAVE_XIII} used a minimum number of circular (and in some cases 
elliptical) two-dimensional Gaussian components that after being convolved with the restoring 
beam, adequately reproduce the constructed brightness distribution of a source. 

\begin{figure*}
\centering
\resizebox{0.32\hsize}{!}{\includegraphics[angle=0]{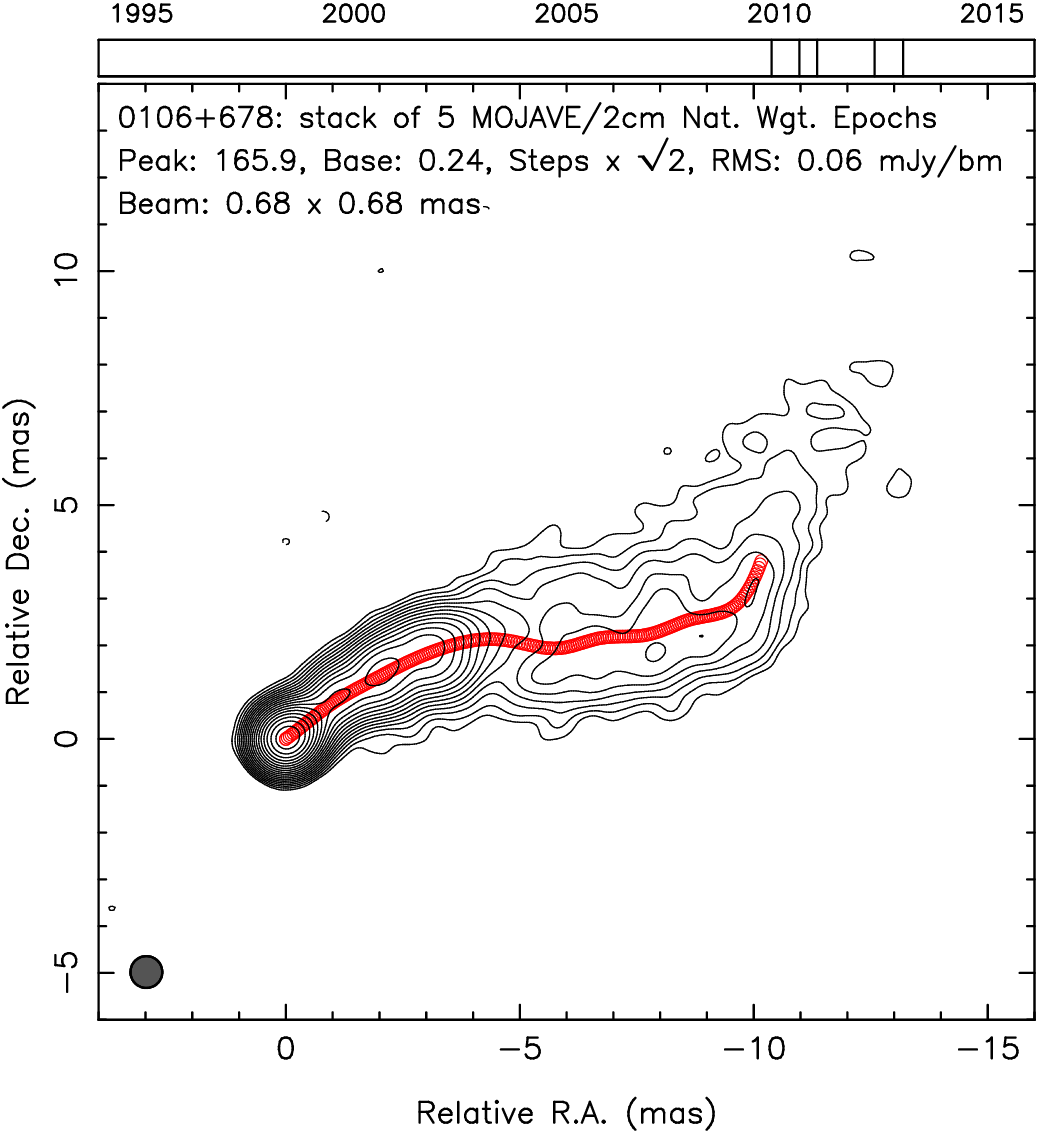}}\hspace{0.1cm}
\resizebox{0.32\hsize}{!}{\includegraphics[angle=0]{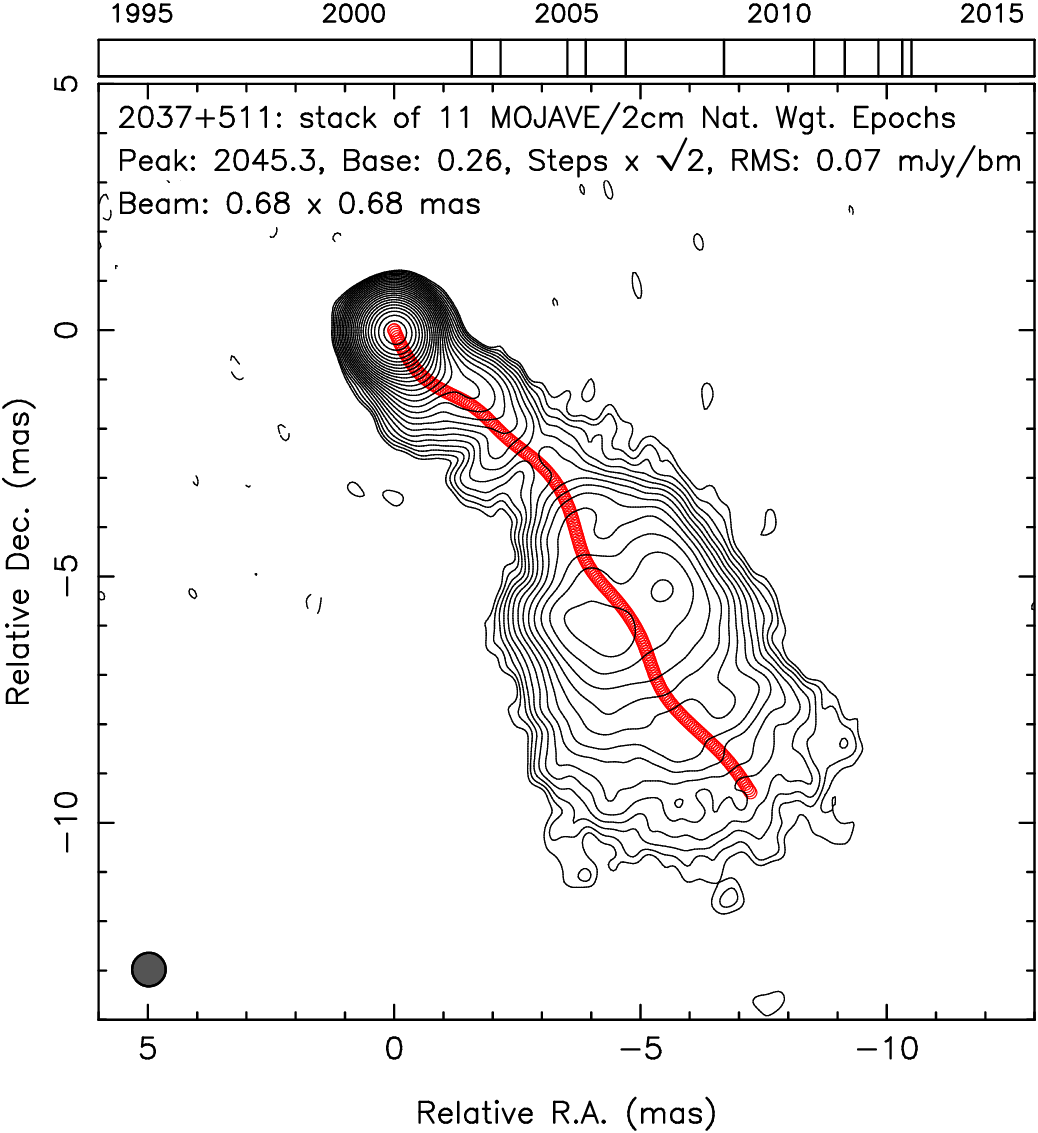}}\hspace{0.1mm}
\resizebox{0.32\hsize}{!}{\includegraphics[angle=0]{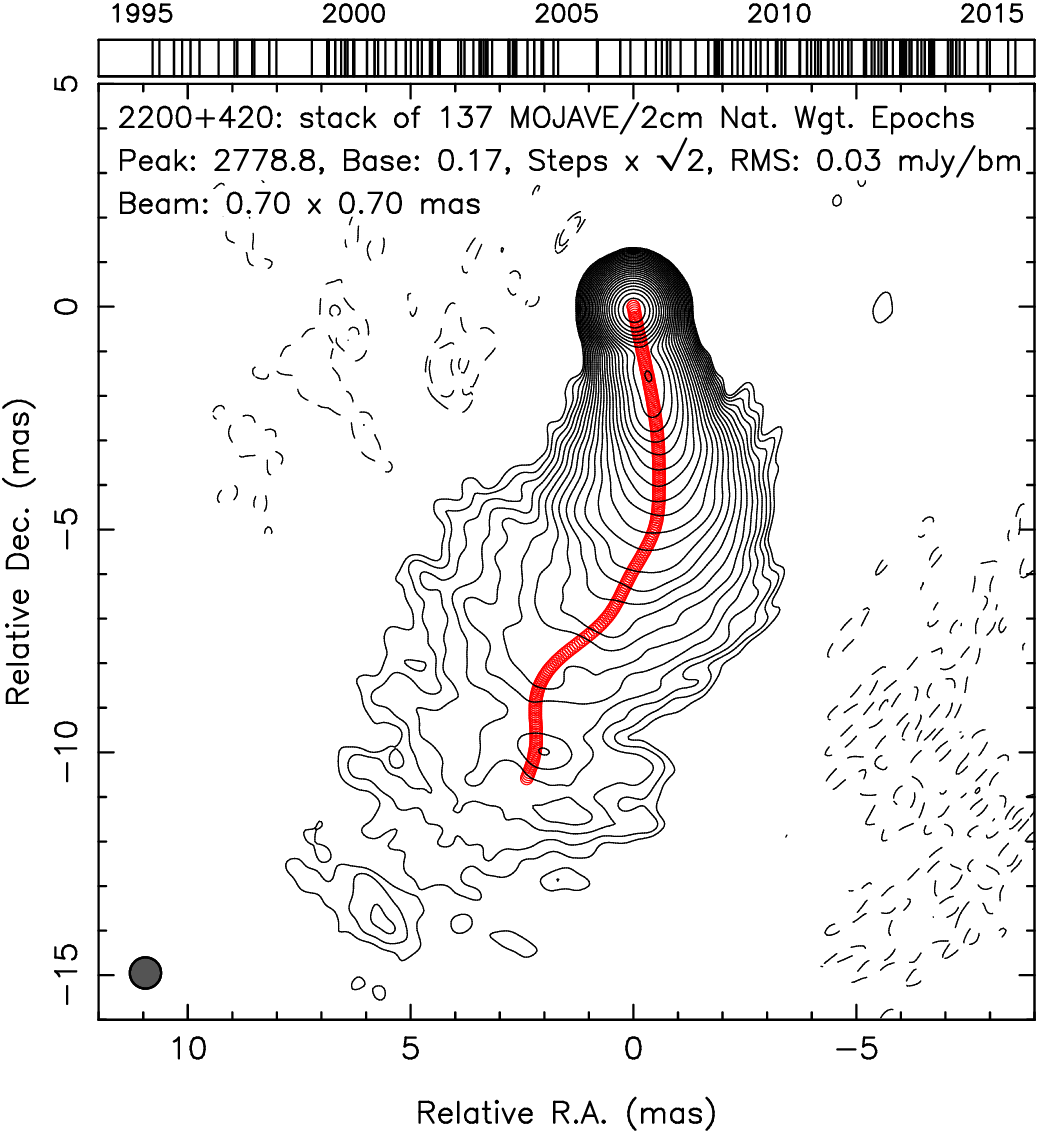}}\vspace{0.1mm}
 \caption{Examples of 15~GHz naturally weighted stacked 
          \textsc{clean} images of observed sources with
          minimum (5), median (11), and maximum (137) number of epochs. The contours are plotted at 
          increasing powers of $\sqrt{2}$. The restoring beam is depicted as a shaded circle in the 
          lower left corner. The constructed total intensity ridgeline is shown by red. A wedge indicating 
          observing epochs (vertical ticks) used for producing the stacked image is shown on top. The 15~GHz 
          VLBA contour stacked images of all 362 sources with constructed ridgelines are available
          in the online journal.}
 \label{f:stacked_maps_examples}
\end{figure*}

The distribution of total observing epochs used in each stacked image is quite broad, ranging 
from the minimum accepted 5 epochs up to 137 epochs (for BL Lac), with a median value of 11. 
The distribution of time range in a stacked map (Fig.~\ref{f:epochs_stat}) shows two humps 
representing two groups of sources dominated by following: (i) gamma-ray bright objects listed as 
high-confidence associations by \textit{Fermi}-LAT after 2008 \citep{3FGL}, and (ii) long-term 
monitored AGNs belonging to both the 2~cm VLBA Survey and MOJAVE programme. The quasar 1928+738 
(4C +73.18) has the longest (20.8~yr) observed time interval in our sample. We list the 
parameters of the stacked images in Table~\ref{t:stacked_map_parameters}. The typical 
FWHM dimension of the circular restoring beam of the stacked VLBA images is about 0.8~mas, 
corresponding to a linear scale of $\sim$6 parsecs in projection at the typical redshifts of 
our sample ($z\simeq1$). The stacking procedure effectively decreases the rms noise of the 
resulting image, on average, by a factor of few compared to that of single-epoch maps. The 
noise level was calculated as a minimum of rms estimates in four corner quadrants of the 
image, each of 1/16 of the image size. For most sources the bottom contour is shown at four 
times the rms level. The dynamic range of the images (determined as a ratio of the peak flux 
density to the rms noise level) ranges from 620 (weak galaxy 0026$+$346) to 92000 (BL Lac) 
with a median of $\sim$11500. The typical rms noise level is $\sim$0.06~mJy~beam$^{-1}$. 
In Fig.~\ref{f:stacked_maps_examples} we show stacked images of three sources having the 
minimum (5), median (11), and maximum (137) number of epochs.

\subsection{22\,cm VLBA data}
In addition to the 2~cm MOJAVE VLBA data, we made use of single-epoch longer wavelength 
VLBA observations of 135 MOJAVE-1 sources that constitute a statistically complete, 
flux-density limited sample. A total of 9 24-hour observing sessions during 2010 were 
carried out by D. Gabuzda et al. in full dual-polarization mode at four wavelengths in 
the range of 18--22~cm at an aggregate recording bit rate of 256~Mbits\,s$^{-1}$. The 
L-band project was originally aimed on studying Faraday rotation properties across the
jet\footnote{\url{http://www.physics.ucc.ie/radiogroup/18-22cm_observations.html}}.
We have processed and imaged the data at the longest wavelength, 22~cm, to reconstruct 
the outer total intensity jet structure and probe large scales that can extend to 100~mas 
or more.

\section{Results}
\label{s:results}
\subsection{Total intensity ridgelines}
\label{ss:ridgelines}

\begin{table*}
\begin{minipage}{156mm}
\caption{Summary of 15 GHz stacked image parameters. Columns are as follows:
(1) B1950 name;
(2) other name;
(3) date of first epoch;
(4) time range between first and last epochs;
(5) number of stacked epochs;
(6) FWHM of restoring beam (milliarcseconds);
(7) I peak of image (Jy per beam);
(8) rms noise level of image (mJy per beam); and
(9) bottom I contour level (mJy per beam).
This table is available in its entirety in the online journal.
A portion is shown here for guidance regarding its form and content.}
\label{t:stacked_map_parameters}
\begin{tabular}{c c c r r c c c c}
\hline
\noalign{\smallskip}
  Source   &                     Alias &  First epoch & $\tau$ &  N  & Beam  & $\text{I}_\text{peak}$ &      rms & $\text{I}_\text{base}$ \\
           &                           & (yyyy/mm/dd) &   (yr) &     & (mas) &                    (Jy/bm) & (mJy/bm) &                   (mJy/bm) \\
     (1)   &                       (2) &          (3) &    (4) & (5) &   (6) &                        (7) &      (8) &                        (9) \\
\hline\noalign{\smallskip}
0003$+$380 &      S4         0003$+$38 &  2006/03/09  &   7.42 &  10 &  0.70 &  0.460 &  0.07 &  0.29 \\
0003$-$066 &    NRAO               005 &  1995/07/28  &  17.27 &  27 &  0.84 &  1.248 &  0.09 &  0.35 \\
0006$+$061 &  CRATES      J0009$+$0628 &  2011/12/29  &   1.43 &   5 &  0.86 &  0.150 &  0.09 &  0.37 \\
0007$+$106 &     III                Zw &  1995/07/28  &  17.85 &  25 &  0.79 &  0.740 &  0.07 &  0.27 \\
0010$+$405 &      4C          $+$40.01 &  2006/04/05  &   5.22 &  12 &  0.70 &  0.507 &  0.03 &  0.11 \\
0011$+$189 &     RGB       J0013$+$191 &  2013/12/15  &   1.42 &   6 &  0.79 &  0.104 &  0.03 &  0.10 \\
0015$-$054 &     PMN      J0017$-$0512 &  2009/07/05  &   4.01 &   8 &  0.83 &  0.231 &  0.04 &  0.17 \\
0016$+$731 &      S5         0016$+$73 &  1994/08/31  &  16.07 &  15 &  0.70 &  0.951 &  0.11 &  0.45 \\
0019$+$058 &     PKS        0019$+$058 &  2014/02/14  &   1.34 &   5 &  0.78 &  0.275 &  0.03 &  0.12 \\
0026$+$346 &      B2         0026$+$34 &  1995/04/07  &   9.24 &   7 &  0.88 &  0.087 &  0.14 &  0.56 \\
\hline
\end{tabular}
\end{minipage}
\end{table*}

\begin{figure*}
\centering
\resizebox{0.32\hsize}{!}{\includegraphics[angle=0]{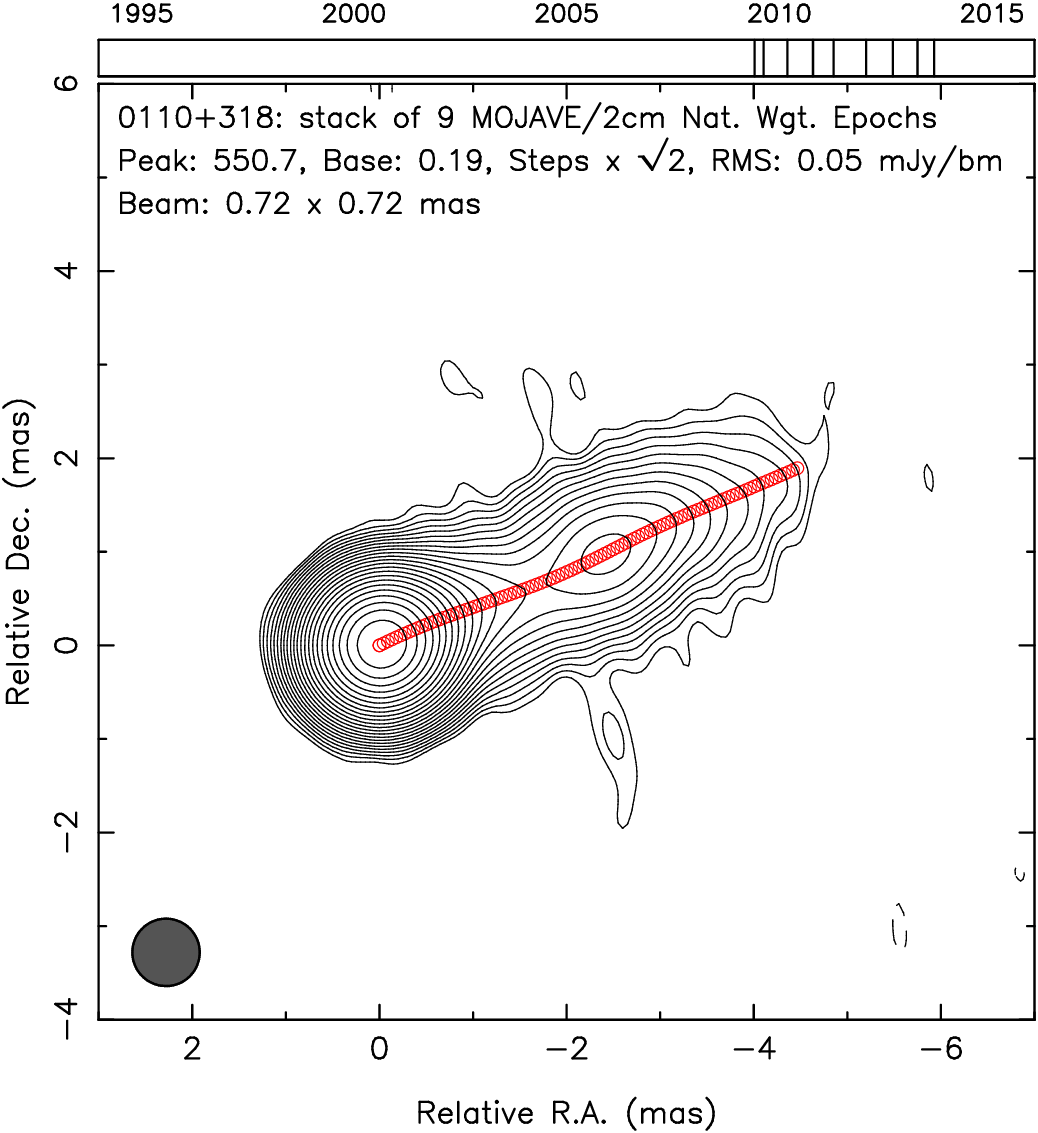}}\hspace{0.5cm}
\resizebox{0.32\hsize}{!}{\includegraphics[angle=0]{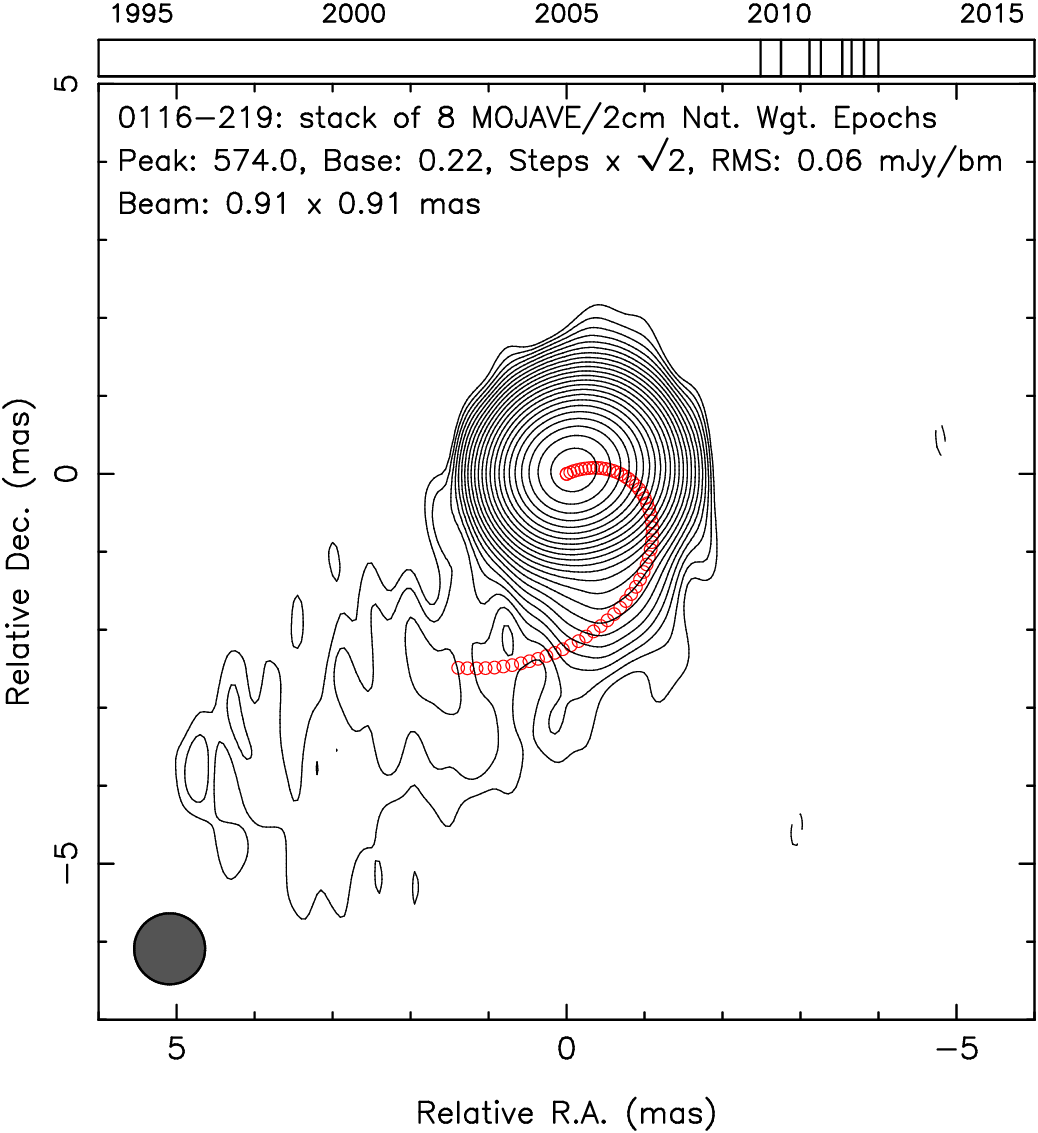}}\vspace{0.1mm}
 \caption{Examples of nearly straight (left-hand panel) and significantly curved (right-hand panel) jets.
          The contours are plotted at increasing powers of $\sqrt{2}$.
          The restoring beam is depicted as a shaded circle in the lower left corner.
          The constructed total intensity ridgeline is shown by red.
          A wedge indicating observing epochs used for producing the stacked image is shown on top.
         }
 \label{f:straight_bend_examples}
\end{figure*}

In our earlier study \citep{Pushkarev09}, we approximated the jet axis by a straight line for
relatively unbroken morphology or by two connected lines for sources with notably bending
outflows. In the current analysis we have constructed total intensity ridgelines to follow
the jet more accurately. The ridge-finding procedure adopts a polar coordinate system centered
on the core and uses an azimuthal slice to find the weighted average, i.e., the point where
the intensity integrated along the arc is equal on the two sides excluding the pixels with
low-SNR ($<8$ times rms). The algorithm advances down the jet for successively increasing
radial values. Finally, the ridgeline is constructed by fitting a cubic spline and interpolating
it at roughly equal intervals (0.05~mas) of radial distance (Fig.~\ref{f:stacked_maps_examples}).
As the  procedure operates in the image plane, an elongated restoring beam could affect the
ridgeline, forcing it to the direction of the major axis. Therefore, all the stacked images were
made from single-epoch maps convolved with a median circular beam. For the sources with two-sided
jet morphology, e.g., NGC~1058, 3C~84, and 1413+135, we set corresponding azimuth limits to
reconstruct the ridgeline of the approaching jet only.

The path lengths along the constructed ridgelines range from 1 to 57~mas, with a median of about 6~mas. 
Due to jet bending being magnified by projection effects, the ridgeline path length is always larger 
than the radial distance from the core to ridgeline final point. The ratio of these values is close to 
1 for straight jets and reaches up to 1.8 for the highly curved outflow of the quasar 0116$-$219, which 
has a continuous change of the ridgeline position angle exceeding $200\degr$ 
(Fig.~\ref{f:straight_bend_examples}). The quasar 2135+141 is known for its extremely bent jet with 
$\Delta\text{PA}=210\degr$ on scales probed by 2--43~GHz VLBI observations \citep{Savolainen06}, and 
shows a total position angle change of about $150\degr$ in the 15~GHz stacked image. On average, the 
standard deviation of jet position angle is inversely proportional to jet length $r$, following the 
median dependence $\sigma_\text{PA[deg]}=7.4/r_\text{[mas]}$.

\subsection{Jet shapes}
\begin{figure}
\includegraphics[width=\columnwidth, angle=0]{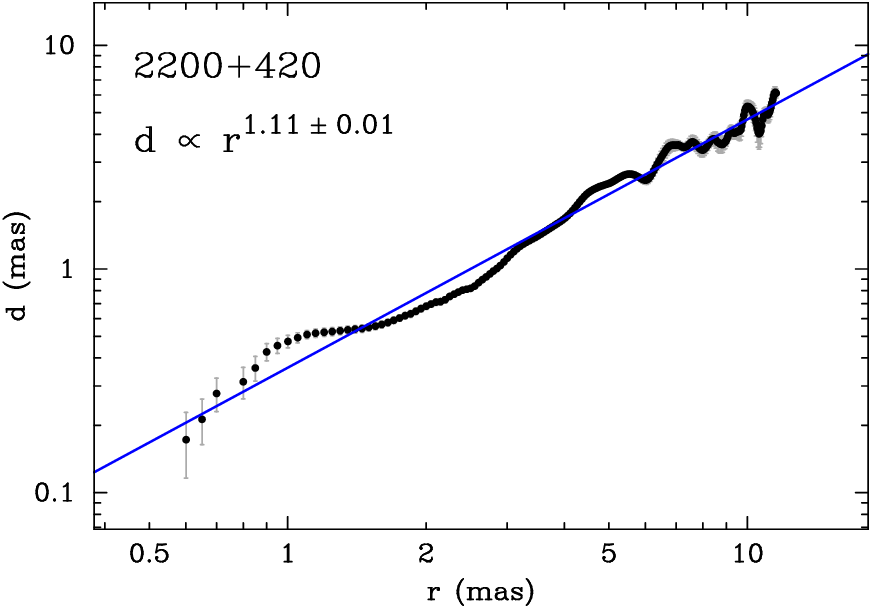}
 \caption{Transverse jet width versus a distance along ridge line for BL Lac at scales probed
          by the MOJAVE observations at 15~GHz. The thick blue line is the best fit of an
          assumed power-law dependence $d\propto r^k$.
          The plots for other sources are available in the online journal.}
 \label{f:bllac_shape}
\end{figure}

We used the 15~GHz total intensity MOJAVE stacked images and constructed ridgelines to study the 
shapes of the outflows. Moving down the ridgeline we made slices transverse to the local jet direction. 
The slices were taken at 0.05~mas intervals along the ridgeline, starting from the position of the core. 
For each cut we obtained the FWHM $D$ of a Gaussian fitted to the transverse jet brightness profile
and the corresponding deconvolved jet width $d=(D^2-b^2)^{1/2}$, where $b$ is the FWHM size of the 
restoring beam. We analysed a dependence between the jet width $d$ and path length $r$ measured along 
the reconstructed total intensity ridge line. We fit an assumed single power-law dependence $d\propto r^k$
using the least squares method through all the distances $r>0.5$~mas. The data points were initially
smoothed by five-point moving averaging. In Fig.~\ref{f:bllac_shape}, we show this dependence for BL Lac,
as an example. The width errors were estimated by deviating a position angle of the transverse jet cut
in a range $\pm15\degr$ with a step of $1\degr$ and calculating the rms.

\begin{figure}
\includegraphics[width=\columnwidth, angle=0]{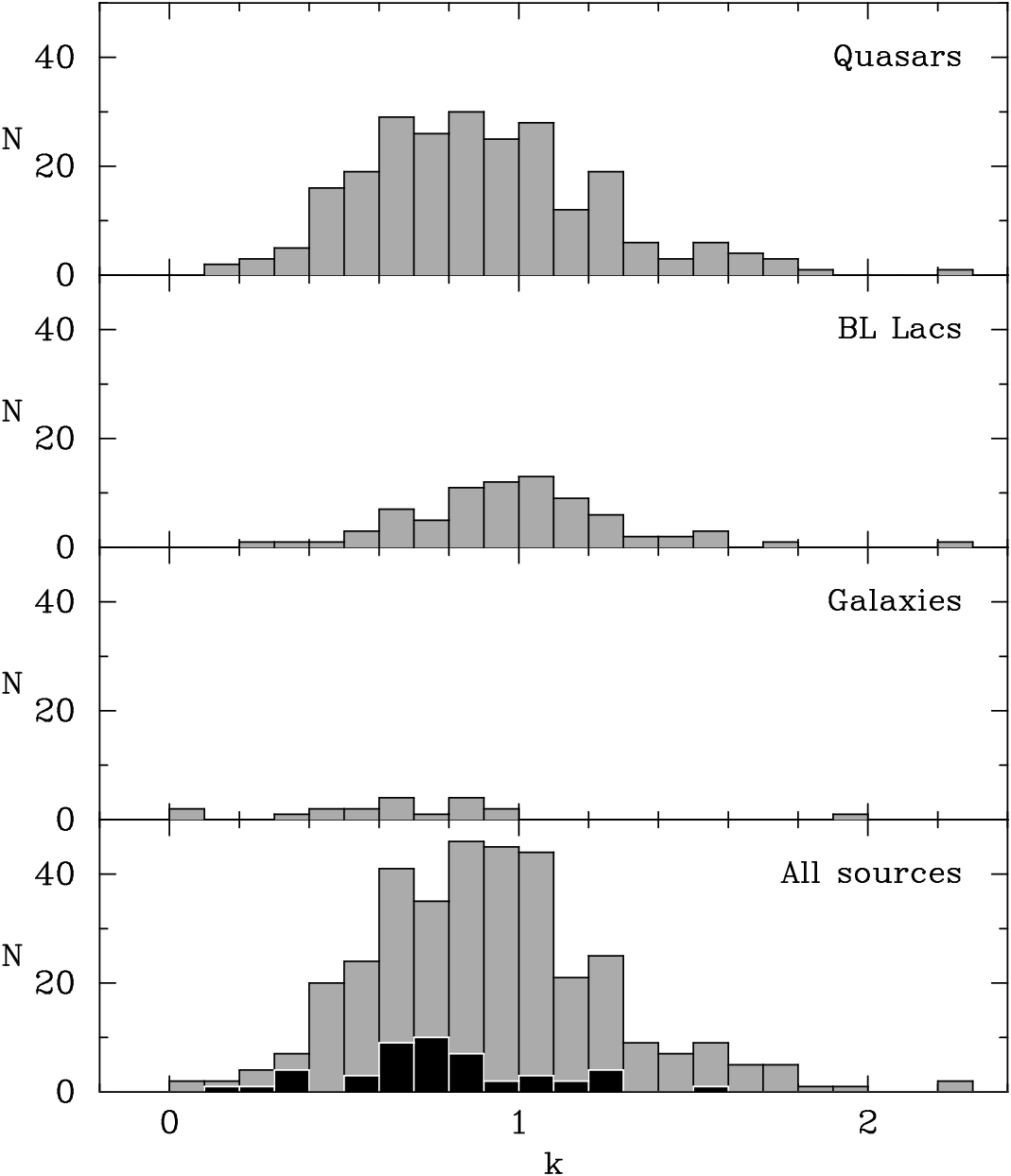}
 \caption{Histograms of the power-law index $k$ in $d\propto r^k$ fitted dependence for
          238 quasars, 78 BL Lacs, 19 galaxies, and all sources together. Black bins
          represent sources with radial accelerated motion detected in their jets.}
 \label{f:collimation_u_band}
\end{figure}

\begin{figure*}
\centering
\resizebox{0.32\hsize}{!}{\includegraphics[angle=0]{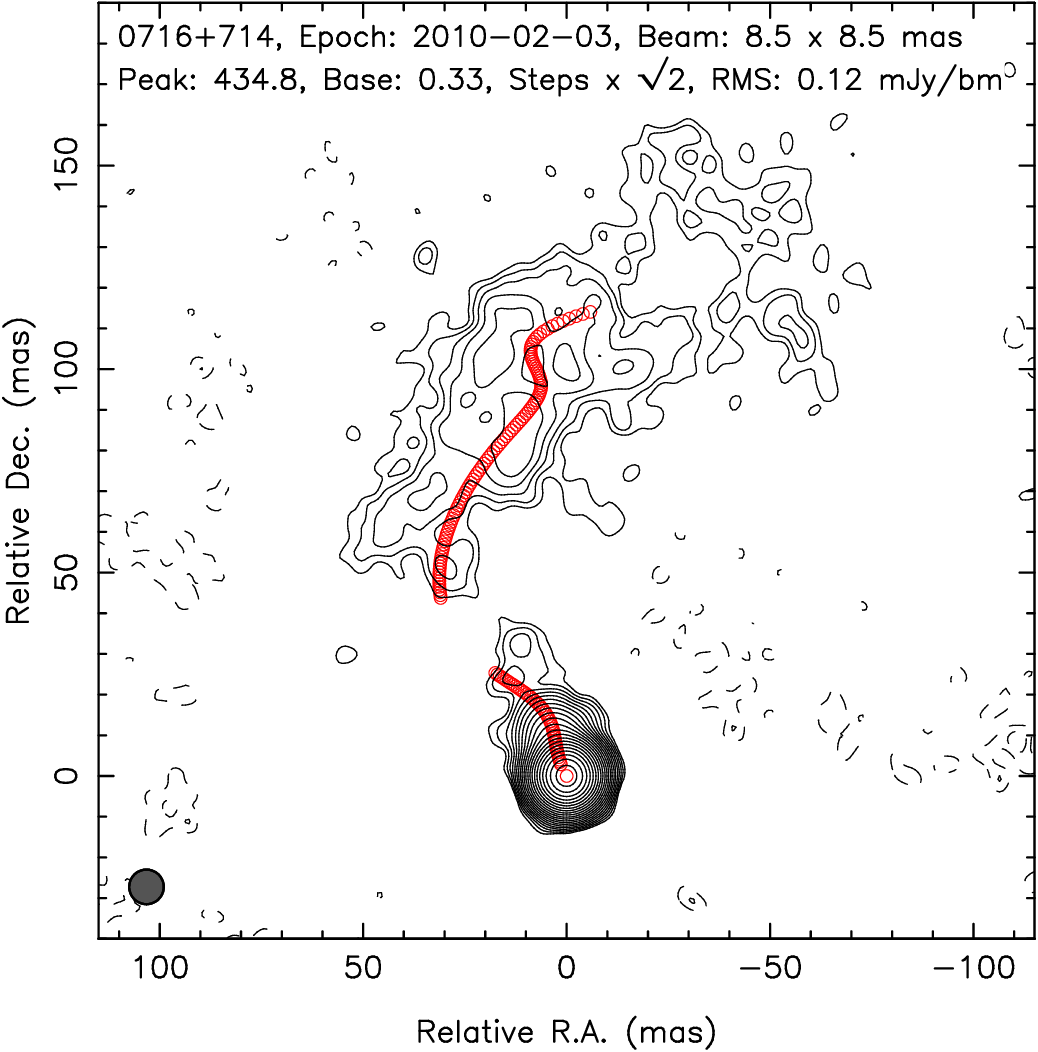}}\hspace{0.1mm}
\resizebox{0.32\hsize}{!}{\includegraphics[angle=0]{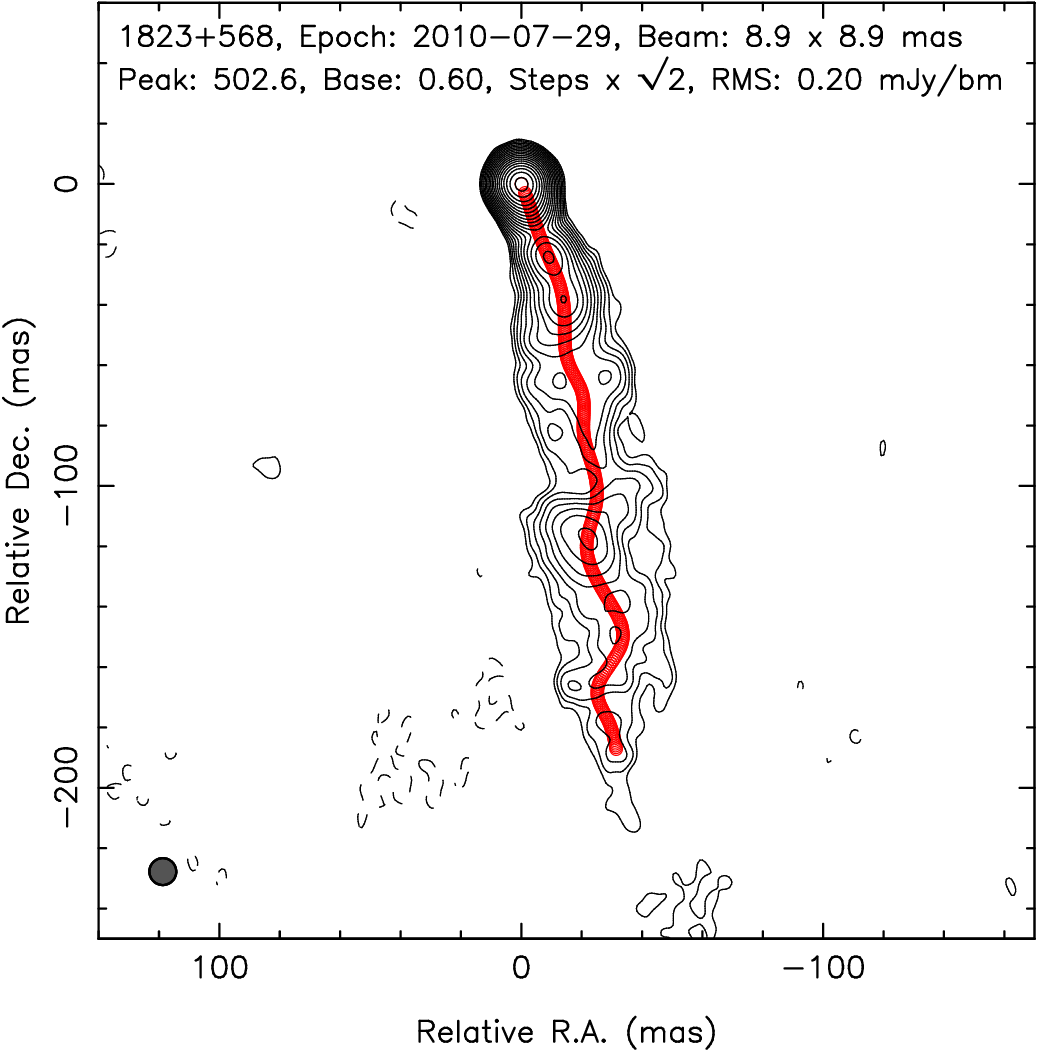}}\hspace{0.1mm}
\resizebox{0.32\hsize}{!}{\includegraphics[angle=0]{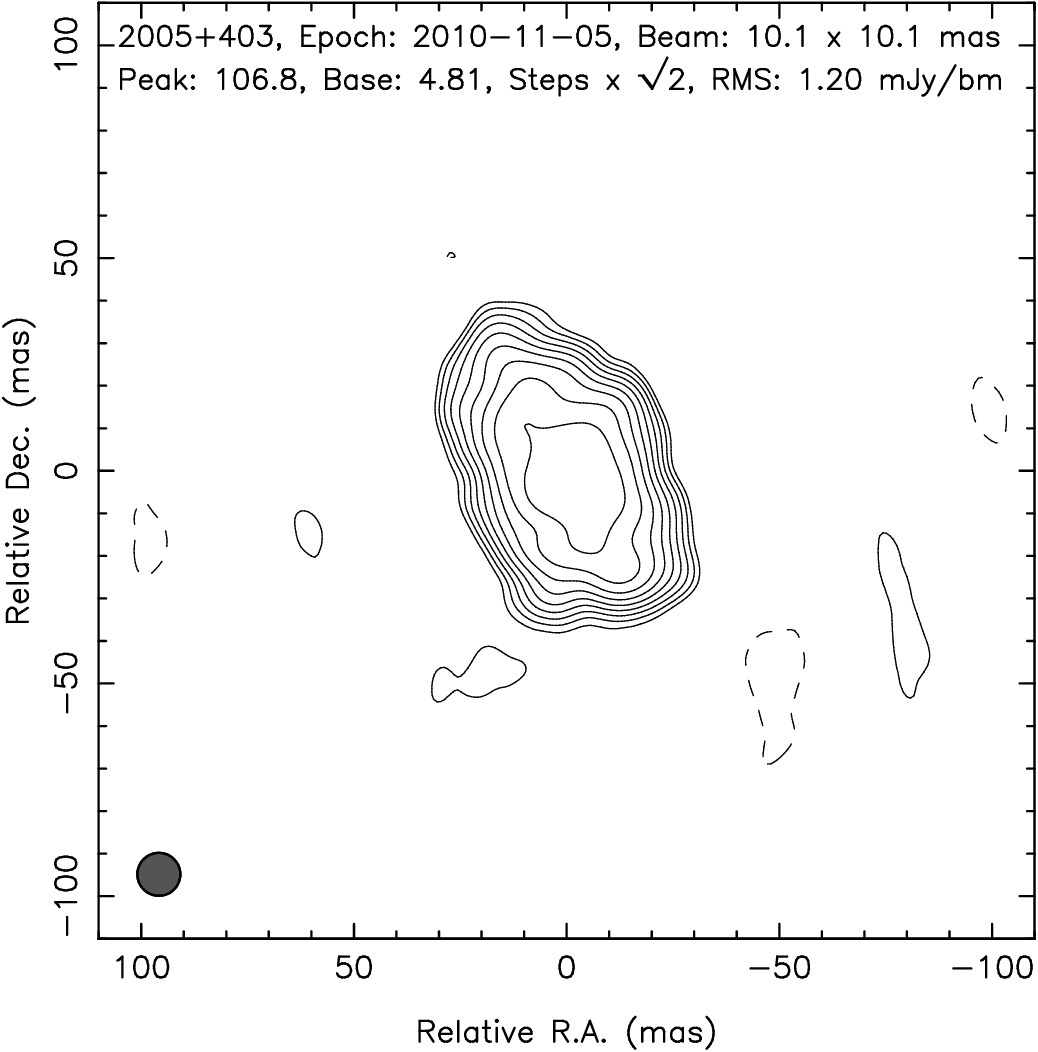}}
 \caption{Examples of 1.4 GHz naturally weighted 
          \textsc{clean} images showing rich jet structure on scales of 100~mas
          or more (left and middle) and a scattered source observed through the highly turbulent Cygnus region.
          The contours are plotted at increasing powers of $\sqrt{2}$.
          The restoring beam is depicted as a shaded circle in the lower left corner.
          The constructed total intensity ridgeline is shown by red.
          The 1.4~GHz VLBA contour images of 122 MOJAVE sources with constructed ridgelines are available
          in the online journal.
         }
 \label{f:maps_lband}
\end{figure*}

Distributions of the power-law index $k$ (Fig.~\ref{f:collimation_u_band}, top three panels) were
constructed separately for sources of different spectral classes. BL Lacs and quasars typically show
jet shape close to conical (i.e. $k=1$), although BL Lacs tend to have on average larger $k$-indices
than those of quasars, with medians of 0.98 and 0.85, respectively. Galaxies are characterized by lower
$k$-index values, with a median of 0.68, suggesting that in a number of cases the jet profile is close
to parabolic geometry, as e.g., in M87 for which $k=0.45$ that is in agreement with other VLBI jet shape
studies of M87 \citep{Asada12,Hada13_M87,Hada16_M87}. Most probably, this is a result of their proximity
and larger viewing angle, allowing us to probe smaller physical scales closer to the jet apex. A jet 
shape transition from parabolic to conical detected in a few sources will be discussed in a separate 
paper (Kovalev et al., in prep.). A dependence between $k$-index and redshift is significant if galaxies 
and BL Lacs are considered only, and becomes non-significant for quasars only. The distribution 
for all sources (Fig.~\ref{f:collimation_u_band}, bottom panel) has a median of 0.89. This is consistent 
with the median $k$-index of 0.8 derived from results of model fitting performed in the $(u,v)$ plane 
for a small sample of 30 sources observed at 8.6~GHz with global VLBI \citep{RDV_paper}.

The overlaid black bins show sources with positive or negative radial accelerated motion observed in 
their jets \citep{MOJAVE_XII}. These objects have statistically a lower $k$ parameter compared to that 
of the rest of sources, with medians of 0.73 and 0.91, respectively, reflecting that the outflows with 
accelerated motion undergo more active collimation.

We tested the robustness of the obtained results by progressively cutting the ridgeline length $r$ down to 1~mas 
from the core and repeating the analysis with the sub-data sets. The corresponding median $k$-index values remain
constant until the ridgeline is up to 10~mas long, while for shorter ridgelines the median gradually increases,
peaking up to 1.0 at $r=1.2$~mas. We explain this by a non-uniform degree of completeness of jet cross-section 
that can decrease beyond a certain distance from the VLBI core. This is because the source brightness distribution 
is stacked over a limited time interval that might be not long enough for jet features emergent at different position 
angles to propagate all the way down to distances where jet emission is detected in the stacked image. At the same 
time, the difference between the accelerated and non-accelerated sources also holds for all the data subsets.
An observational bias that can also affect the $k$-index measurements to some degree is that there is a maximum 
angular scale to which the VLBA is sensitive. It is about $0.5\lambda/B_\text{min}$, where $\lambda$ is the 
wavelength of observation and $B_\text{min}\approx236$~km is the minimum baseline length (between the antennas Los 
Alamos and Pie Town, New Mexico). This corresponds to the scale of about 9~mas at 2~cm, while about 75\% of sources 
in our sample manifest jet structure on smaller scales, making this bias weak.

\begin{table*}
\begin{minipage}{129mm}
\caption{Summary of 1.4~GHz single-epoch image parameters. Columns are as follows:
(1) B1950 name;
(2) other name;
(3) observing epoch;
(4) FWHM of restoring beam (milliarcseconds);
(5) I peak of image (Jy per beam);
(6) rms noise level of image (mJy per beam); and
(7) bottom I contour level (mJy per beam).
This table is available in its entirety in the online journal.
A portion is shown here for guidance regarding its form and content.}
\label{t:lband_map_parameters}
\begin{tabular}{c c c r r c c c c}
\hline
\noalign{\smallskip}
Source   &                   Alias &        Epoch &  Beam  & $\mathrm{I}_\mathrm{peak}$ &      rms & $\mathrm{I}_\mathrm{base}$ \\
         &                         & (yyyy/mm/dd) &  (mas) &                    (Jy/bm) & (mJy/bm) &                   (mJy/bm) \\
   (1)   &                     (2) &          (3) &    (4) &                        (5) &      (6) &                        (7) \\
\hline\noalign{\smallskip}
0003$-$066 &    NRAO           005 &  2010/11/05  &  14.4  &  1.983 &  0.48 &  1.90 \\
0016$+$731 &      S5     0016$+$73 &  2010/11/05  &   8.7  &  0.314 &  0.11 &  0.44 \\
0048$-$097 &     PKS     0048$-$09 &  2010/06/18  &  13.5  &  0.218 &  0.13 &  0.50 \\
0059$+$581 &     TXS    0059$+$581 &  2010/07/29  &   9.5  &  0.682 &  0.18 &  0.73 \\
0106$+$013 &      4C      $+$01.02 &  2010/08/23  &  13.4  &  1.445 &  0.36 &  1.44 \\
0109$+$224 &      S2     0109$+$22 &  2010/09/23  &  11.6  &  0.209 &  0.09 &  0.35 \\
0119$+$115 &     PKS     0119$+$11 &  2010/05/21  &  12.1  &  1.102 &  0.22 &  0.88 \\
0133$+$476 &      DA            55 &  2011/08/18  &   9.5  &  1.127 &  0.29 &  1.16 \\
0202$+$319 &      B2     0202$+$31 &  2010/03/07  &  10.2  &  0.542 &  0.16 &  0.63 \\
\hline
\end{tabular}
\end{minipage}
\end{table*}

\subsubsection{Larger scales at 1.4 GHz}
Due to a steep spectrum of the jet synchrotron emission, with a typical spectral index $-0.7$ 
measured between 2 and 8~GHz \citep{RDV_paper} and $-1.0$ between 8 and 15~GHz \citep{MOJAVE_XI}, 
observations at lower frequencies probe larger scales of the outflows, can be used to effectively 
reconstruct the jet cross-section even with single-epoch data. Using the VLBA observations at 
1.4~GHz, we imaged jet structure on scales roughly one order of magnitude larger than those at 
15~GHz, extending to 100~mas or more (Fig.~\ref{f:maps_lband}, left, middle). Typical dynamic 
range of the images is about 3900, with a median noise level of about 0.3~mJy/beam. We list the 
parameters of the single-epoch 1.4~GHz total intensity images in Table~\ref{t:lband_map_parameters}. 
Out of 135 observed sources, we constructed total intensity ridgelines using the procedure described 
in Sect.~\ref{ss:ridgelines} for 122 with a clear core-jet morphology, omitting the sources that 
are either too compact or have unclear core position. One source, the quasar 2005+403, seen through 
the turbulent Cygnus region, is found to be heavily scattered (Fig.~\ref{f:maps_lband}, right). 

\begin{figure*}
\centering
\resizebox{0.42\hsize}{!}{\includegraphics[angle=0]{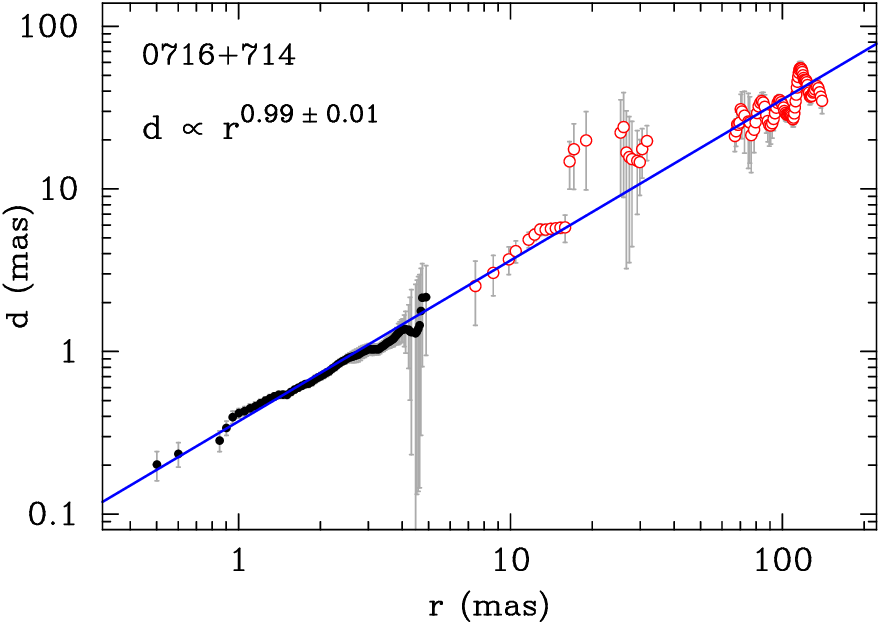}}\hspace{10.1mm}
\resizebox{0.42\hsize}{!}{\includegraphics[angle=0]{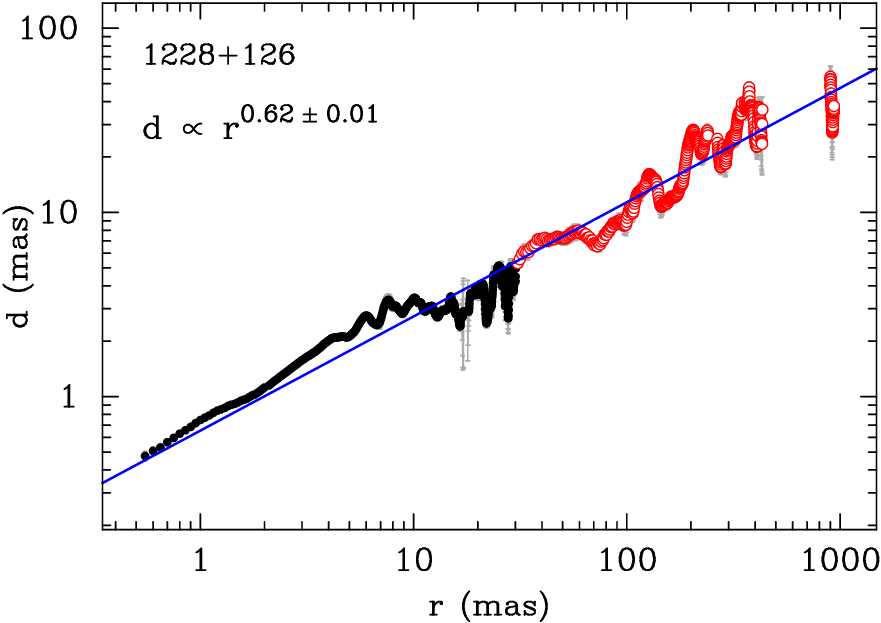}}
 \caption{Transverse jet width versus a distance along ridge line for the BL Lac object 0716+714 
          (left panel) and radio galaxy M87 (right panel) showing conical and parabolic jet shape, 
          respectively, at scales probed by the observations at 15 (filled black dots) and 1.4~GHz 
          (open red circles). The solid blue line is the best fit of an assumed power-law dependence 
          $d\propto r^k$. The plots for other sources are available in the online journal.}
 \label{f:collimation_ul_bands}
\end{figure*}

\begin{figure}
\includegraphics[width=\columnwidth, angle=0]{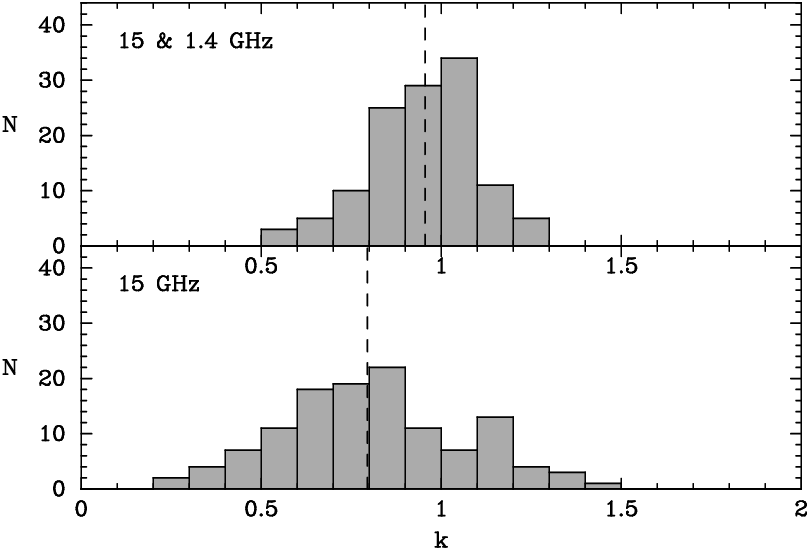}
 \caption{Histograms of the power-law index $k$ in $d\propto r^k$ fitted dependence for 122 MOJAVE-1
          sources derived from the 15~GHz data only (bottom) and from combined 15 and 1.4~GHz
          measurements (top), with medians of 0.80 and 0.95, respectively, shown by dashed lines.}
 \label{f:collimation_ul_band_MOJAVE1}
\end{figure}

Another aspect of the low-frequency observations is that the absolute position of the VLBI core
at 1.4~GHz is expected to be shifted down the jet with respect to the 15~GHz core position mainly
due to synchrotron self-absorption \citep[e.g.,][]{Lobanov_98,Sokolovsky_11,Kutkin14,Kravchenko16}. 
The magnitude of the core shift effect between 15 and 1.4~GHz can be estimated using a typical 
value of 0.13~mas derived statistically between 15 and 8~GHz \citep{MOJAVE_IX} or 0.44~mas between 
8 and 2~GHz \citep{Kovalev_08_cs} and assuming that a separation of the VLBI core from the jet apex 
varies with frequency as $r_\text{core}\propto\nu^{-1}$. This yields a shift of the order of 1~mas, 
which is consistent with typical core shift 1.2~mas between 1.4 and 15~GHz measured by 
\cite{Sokolovsky_11} for a sample of 20 sources. The core shifts for individual sources in our 
sample can not be derived from the single-epoch 1.4~GHz images and 15~GHz stacked maps since the 
stacking procedure smooths the temporally evolving source brightness distribution. Moreover, the 
core shift can vary significantly, especially during flares (Plavin et al, in prep.). Therefore, 
we combined the high and low-frequency $(r,d)$ measurements applying no relative shift in $r$, 
but restricting the 1.4~GHz data sets to the distances farther away from the core not covered by 
the 15~GHz observations. At these outer scales the core shift effect is negligible for the fitting 
algorithm as it works in a logarithmic scale. In Fig.~\ref{f:collimation_ul_bands}, we show the 
combined data sets and corresponding fits. Typically, the truncated 1.4~GHz data at larger scales 
follow the trend of the 15~GHz measurements quite well, without any essential break in the slope. 
The BL Lac object 0716+714 (Fig.~\ref{f:collimation_ul_bands}, left) shows a conical jet geometry 
at scales up to 120~mas, corresponding to about 3~kpc deprojected distance. The radio galaxy M87 
(Fig.~\ref{f:collimation_ul_bands}, right) manifests a parabolic jet profile with a power-law index 
$k$ of 0.62 up to scales of 900~mas, where the HST-1 feature is detected, corresponding to deprojected 
linear distance of 150~pc, assuming viewing angle $30\degr$ \citep{Hada16_M87}. A distribution of 
$k$-index derived from the combined data for the 122 MOJAVE-1 sources 
(Fig.~\ref{f:collimation_ul_band_MOJAVE1}, top) is narrower and shifted to higher values comparing 
to those calculated from the 15~GHz data only (Fig.~\ref{f:collimation_ul_band_MOJAVE1}, bottom). 
For a subsample of 61 sources with inferred viewing angles, known redshifts, and measured projected 
jet lengths, we calculated the linear deprojected jet lengths, the median of which is about 6~kpc. 
Thus, the jet shape of the majority of sources from the MOJAVE-1 sample is still close to conical 
on scales up to a few kpc. We list the jet collimation parameters in Table~\ref{t:inferred_parameters}.

An alternative approach to studying jet shape based on measuring VLBI core size and frequency-dependent 
shift of the core position was successfully used by \cite{Nakamura13_M87} and \cite{Hada13_M87} to 
investigate the outflow of M87. Application of this method by \cite{Algaba16} for a sample of 56 
radio-loud AGNs observed non-simultaneously in a wide frequency range from 1.6 to 86~GHz showed a 
diversity of jet geometries, with a typical streamline close to conical.

As it was recently shown by \cite{Kovalev17}, many AGNs should exhibit prominent parsec-scale jets 
in the optical band. Since the synchrotron opacity is extremely low at these frequencies, an ability 
to optically image jets at high angular resolution to probe the jet geometry in the true jet base.

We also note that interstellar scattering resulting in angular broadening is not substantial at 15~GHz 
\citep{Pushkarev15} and therefore should not affect our results. The 1.4~GHz images can be affected by 
scattering, especially for the sources seen through the Galactic plane, as in the quasar 2005+403 
(Fig.~\ref{f:maps_lband}, right). Effects responsible for a severe distortion of the VLBA image of an 
AGN jet are extremely rare, and have been found by us at 15~GHz for one source only, the quasar 2023+335 
\citep{Pushkarev13}, which is excluded from the analysis of this paper.

\subsection{Apparent opening angles}
\begin{figure}
\centering
\includegraphics[angle=-90,width=0.95\columnwidth]{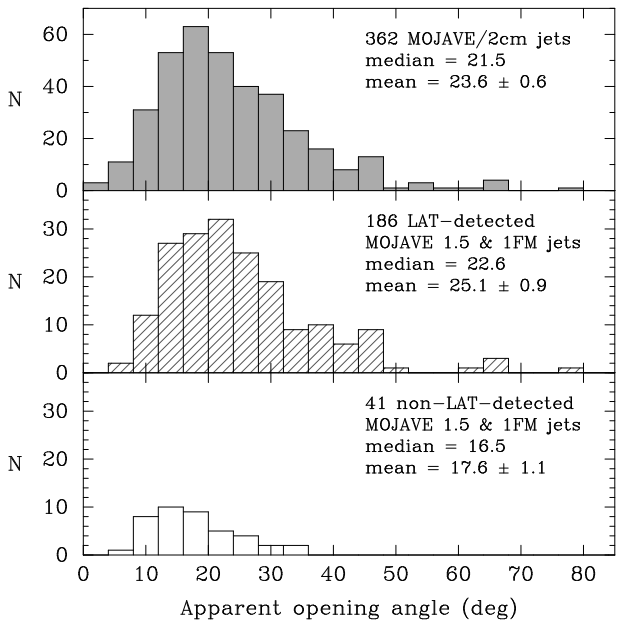}
\caption{Distributions of the apparent opening angle from jet-cut analysis for all 362 MOJAVE AGNs
         (top panel), 186 \textit{Fermi} LAT-detected (middle panel) and 41 non-LAT-detected 
         (bottom panel) sources from the MOJAVE 1.5 Jy and 1FM samples.
\label{f:aoa_LAT_YN}
}
\end{figure}

Having measured deconvolved transverse jet widths $d$ at different path lengths $r$ along the ridge line,
we calculated the apparent opening angle of the jet as the median value of 
$\alpha_\text{app}=2\arctan(0.5d/r)$ for $r>0.5$~mas, where $d=(D^2-b^2)^{1/2}$ is the deconvolved 
FWHM transverse size of the jet, and $b$ is the FWHM size of the restoring beam.
In Fig.~\ref{f:aoa_LAT_YN} (top panel) we show a histogram of the derived projected opening angles for 
all 362 sources. The distribution is broad, comprising values from a few degrees up to $77\degr$ for 
the gamma-ray bright quasar 1520+319, with the median $\alpha_\text{app}=21\fdg5$. Distributions
of $\alpha_\text{app}$ for BL Lacs and quasars are statistically indistinguishable, with similar medians 
of $22\fdg4$ and $21\fdg1$, respectively. Galaxies show significantly narrower apparent opening 
angles of their outflows, with a median of $9\fdg6$, likely due to their statistically larger viewing 
angles. For the 122 MOJAVE-1 sources observed at 1.4~GHz and having constructed ridgelines we measured 
transverse widths and derived apparent opening angles. The latter agree well with $\alpha_\text{app}$ 
at 15~GHz, but on average are about 10\% wider.

Blazar samples are biased in multiple ways \citep[e.g.,][]{Vermeulen94} but using well-defined, complete 
samples can eliminate many of them. Therefore, to make a proper comparison of $\alpha_\text{app}$ 
between LAT-detected and non-LAT-detected jets, we used only the sources from the MOJAVE 1.5 Jy 
\citep{Lister15} and 1FM samples \citep{MOJAVE_X}, representing the brightest radio and gamma-ray AGN 
jets in the northern sky, respectively. The radio-selected MOJAVE 1.5 Jy sample includes all AGNs 
(excluding gravitational lenses) with J2000 declination $>-30\degr$ and VLBA flux density 
$S_\text{15\,\,GHz}>1.5$~Jy at any epoch between 1994.0 and 2010.0. The gamma-ray selected 1FM sample 
based on the initial 11 month First \textit{Fermi} AGN catalog \citep{1LAC} includes all AGNs with 
an average integrated $>0.1$~GeV energy flux $>3\times10^{-11}$~erg~cm$^{-2}$~s$^{-1}$ and J2000 declination 
$>-30\degr$. Out of 227 sources with measured $\alpha_\text{app}$ from these two samples, 186 (82\%) 
objects are positionally associated with the gamma-ray bright sources detected by the \textit{Fermi}-LAT 
during the first 48 months of survey data and included in the third full catalog \citep[3FGL;][]{3FGL} 
or earlier releases \citep{1FGL,2FGL}. The LAT-detected sources (Fig.~\ref{f:aoa_LAT_YN}, middle panel) 
have statistically wider apparent jet opening angles compared to those of non-LAT-detected 
(Fig.~\ref{f:aoa_LAT_YN}, bottom panel). A Kolmogorov-Smirnov (K-S) test indicates a probability $p=0.002$  
($p=0.017$ for 177 sources from the MOJAVE 1.5 Jy sample only), implying that the null hypothesis of the LAT 
and non-LAT sub-samples coming from the same parent population is rejected. This confirms our earlier findings 
\citep{Pushkarev09} at a much higher level of significance as a result of stronger statistics both in the 
radio and gamma-ray domains. The LAT-detected AGN jets of the southern sky studied by the TANAMI programme
were also found to have wider apparent jet opening angles than non-LAT-detected \citep{Ojha10}.

We note that all sources with an apparent jet opening angle wider than $35\degr$ are LAT-detected. The 
apparent jet opening angles of the 105 brightest gamma-ray jets (1FM) are found to be wider then 
those of the 177 brightest radio jets (MOJAVE 1.5~Jy), with medians of $24\fdg1$ and $20\fdg2$, 
respectively. The corresponding  distributions are significantly different ($p_\text{K-S}=0.026$).

Our approach of assessing an apparent jet opening angle assumes a conical jet shape, which is not the case
for a number of sources (Fig.~\ref{f:collimation_u_band}), most noticeably the radio galaxies of the sample, 
and to some extent also a significant number of quasars, that manifest jet profiles close to parabolic, 
implying a decrease of $\alpha_\text{app}$ down the jet on scales probed by our observations. For such sources, 
our method provides an intermediate value of $\alpha_\text{app}$, and which may introduce a bias. Therefore, 
to test this we restricted the calculations to those sources that present a close to conical shape, with 
$0.7<k<1.3$. The obtained distributions and their parameters are close to those from the samples with no 
jet shape restriction, with an even stronger difference between distributions of $\alpha_\text{app}$ for 
the LAT-detected and non-LAT-detected sources ($p=0.0017$).

\begin{figure}
 \includegraphics[width=\columnwidth, angle=0]{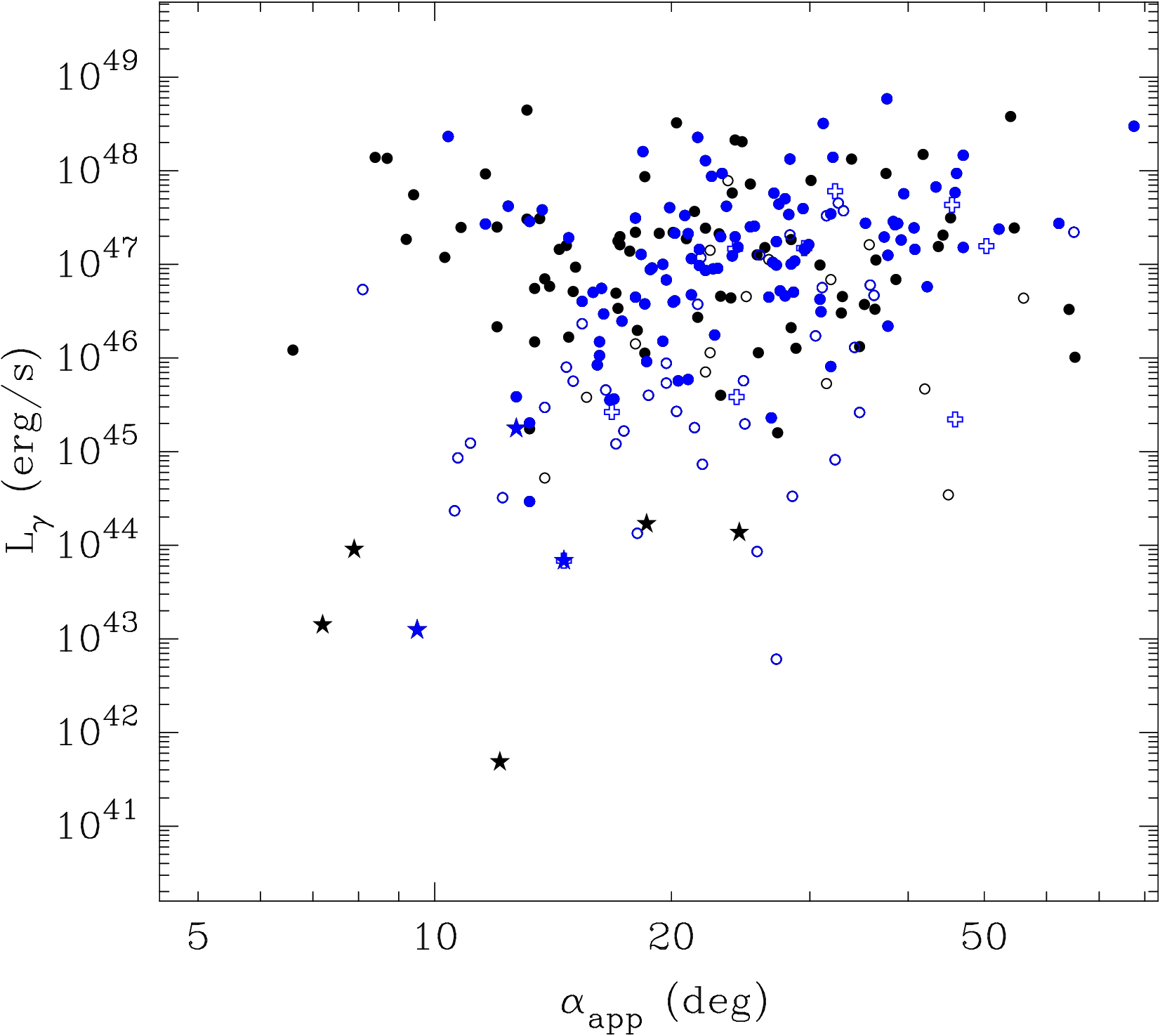}
 \caption{Highly significant correlation between gamma-ray luminosity and apparent jet opening
          angle for the 250 LAT-detected sources. Filled and empty circles show quasars and BL Lacs,
          respectively, star symbols denote galaxies, and empty crosses represent optically
          unidentified sources. The sources with a quasi-conical jet shape are shown by blue.}
 \label{f:gamma_lum_aoa}
\end{figure}

We have also established a highly significant correlation (Kendall's $\tau=0.2$, $p=1.6\times10^{-5}$), 
most probably driven by Doppler beaming, between apparent opening angle and gamma-ray luminosity 
(Fig.~\ref{f:gamma_lum_aoa}) derived from 0.1--300~GeV energy fluxes for 250 sources. The correlation 
remains significant also for different optical classes considered separately. If a sub-sample of sources
characterized by quasi-conical jet shapes is considered (Fig.~\ref{f:gamma_lum_aoa}, blue color), 
the correlation becomes even stronger ($\tau=0.3$; $p=10^{-8}$). These findings suggest that the 
gamma-ray bright AGN jets have on average smaller viewing angles (see more detailed analysis in 
Sec.~\ref{s:ioa_va}).

\begin{figure}
%\vspace{5.9cm}
\includegraphics[width=\columnwidth, angle=0]{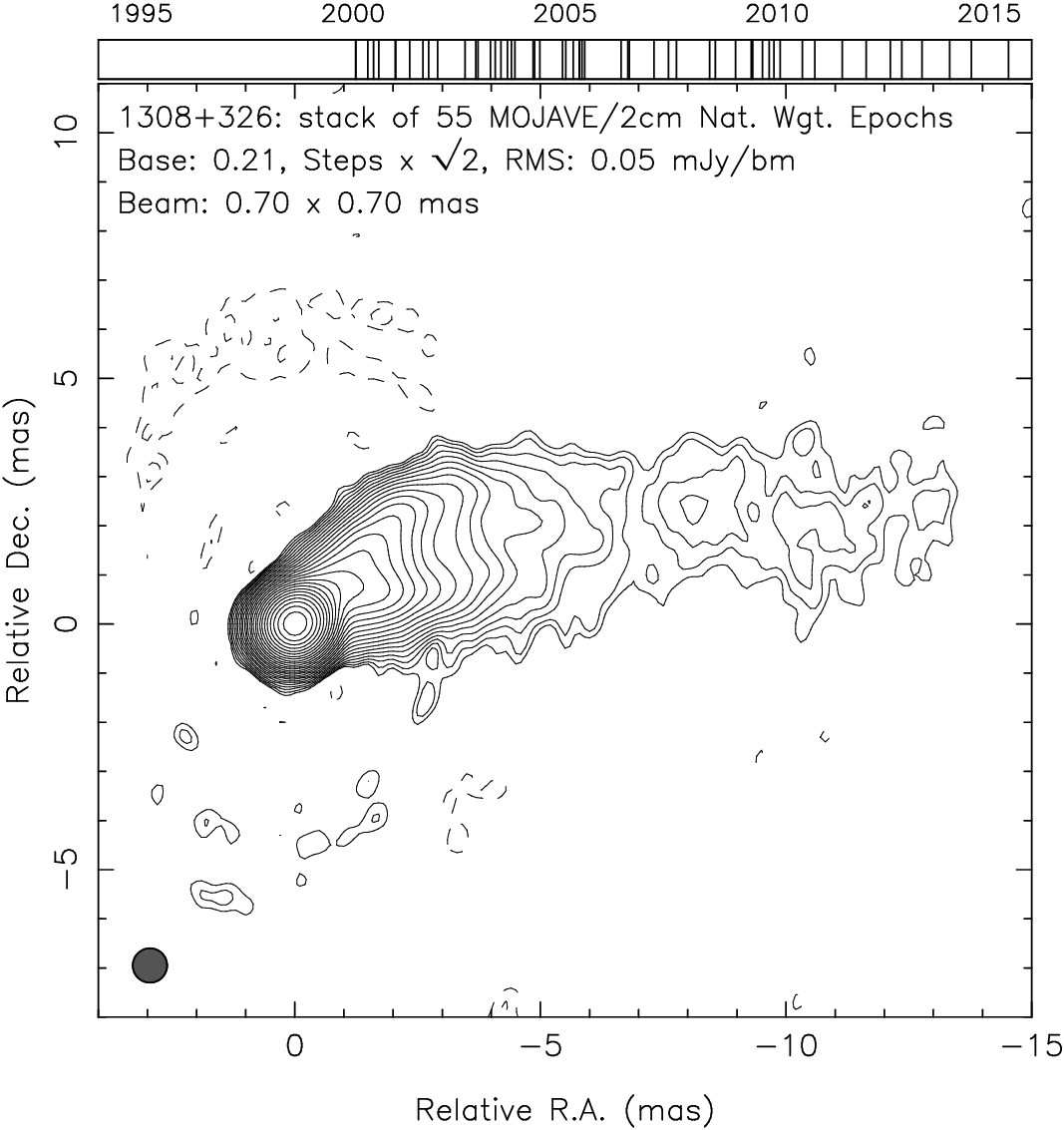}
\caption{Animated evolution showing the build-up of the stacked image of BL Lac 
         object 1308+326 since the epoch of Jan 22, 2000, continuously adding 54 later epochs until 
         16 June, 2015.}
 \label{f:1308+326_animation}
\end{figure}

\begin{figure}
 \includegraphics[width=\columnwidth, angle=0]{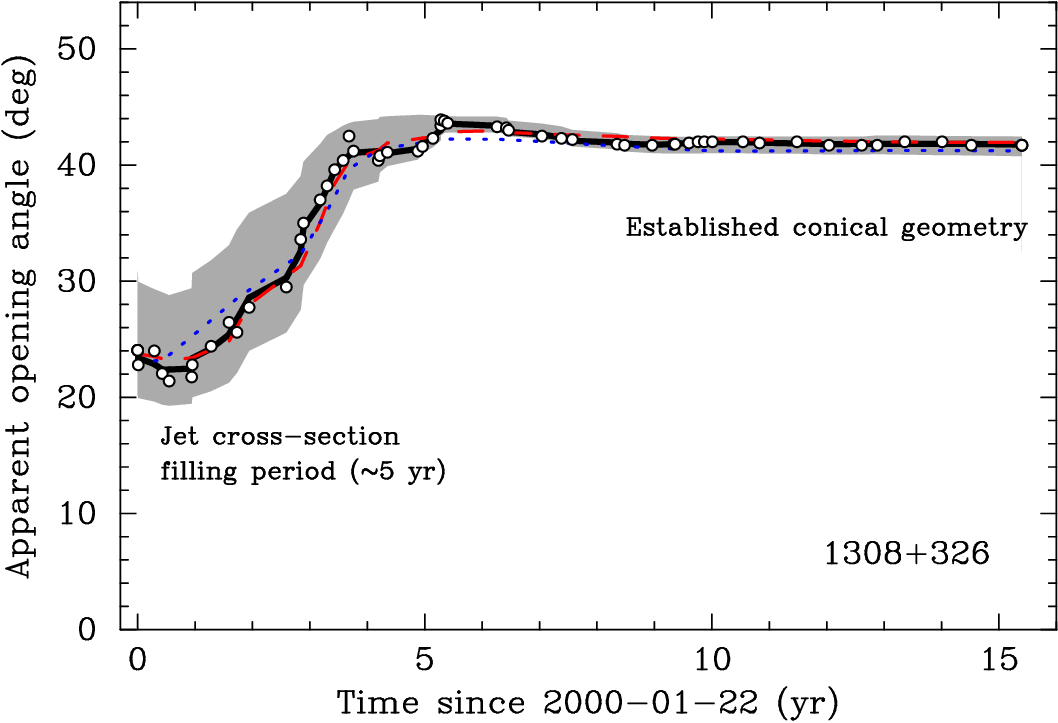}
 \caption{Apparent jet opening angle of BL Lac object 1308+326 as a function of time
          range in a series of stacked images made with progressively increasing number of 
          epochs (up to 55) since Jan 22, 2000. The thick curve is constructed by applying 
          a five-point moving average. 
          Grey area indicates the $1\sigma$ error level of median 
          apparent opening angles, calculated using bootstrap approach. The dashed red and 
          dotted blue lines represent cases with every second and every third epoch used 
          in stacking, respectively.}
 \label{f:1308+326_aoa}
\end{figure}

As the variation of parsec-scale jet orientation is found to be a common phenomenon for AGNs observed 
on decadal time-scales \citep{MOJAVE_X}, we studied how this effect influences the measured apparent 
jet opening angle and how the latter evolves with adding more epochs to the stacked image of a source. 
For this analysis we used the data of BL Lac object 1308+326, which is one of 12 MOJAVE sources that 
manifested oscillatory trends of the innermost jet position angle as reported by \cite{MOJAVE_X}.
This source did not display a pc-scale jet in the early 1990's and the opening angle was thus undefined. 
We therefore used data since the epoch of Jan 22, 2000, after which the jet was well-detected. We made a 
series of stacked images of the source, continuously adding 54 later epochs until 16 June, 2015 inclusive, 
and measured the apparent opening angle following the procedure described above. The animation of stacked 
image evolution demonstrates that the jet geometry eventually becomes close to conical 
(Fig.~\ref{f:1308+326_animation}). The corresponding evolution of $\alpha_\text{app}$ is shown in 
Fig.~\ref{f:1308+326_aoa}. The apparent opening angle was of order of $23\degr$ until the end of 2000 
and then widened to about $40\degr$ during about five next years due to emergence of a new bright jet 
feature in a different position angle, and then $\alpha_\text{app}$ remains constant. The plateau of 
$\alpha_\text{app}$ means that the jet cross-section is effectively filled out by stacking. We performed 
the same analysis with lower time sampling by using (i) every second and (ii) every third epoch out of 
the 55 available. In both cases, the obtained dependence of $\alpha_\text{app}$ versus time range covered 
by stacking is very close to the original. This implies that the time period of about 5~yr is not an 
observation specific bias, but rather a source specific characteristic.

Of course, the duration of the jet cross-section filling period is expected to be source dependent. 
To place constraints on this we studied the dependence between median jet width $d_\text{med}$ and 
a number of epochs $N$ in a stacked image for different sets formed by changing minimum time interval 
$\Delta t$ covered by stacking, and analysed the corresponding Kendall's correlation statistics. The 
dependence becomes significant ($p<0.05$) at $\Delta t\ga2$~yr, and it is strongest at the minimum 
time coverage of $\sim$6~yr. This implies that an average source changes the inner jet position 
angle on a time-scale of at least 6~yr. On the other hand, for sources having more than 30 epochs, 
no significant dependence between $d_\text{med}$ and $N$ is detected. This corresponds to the time
range $\Delta t\ga16$~yr, setting upper limit on position angle variation time-scale for majority
of sources in our sample. This is consistent with indications of oscillatory behaviour of jet
orientation with best fitting periods ranging from 5 to 12~yr found by \cite{MOJAVE_X}.

\subsection{Intrinsic opening angles and viewing angles}
\label{s:ioa_va}
\begin{figure}
 \includegraphics[width=\columnwidth, angle=0]{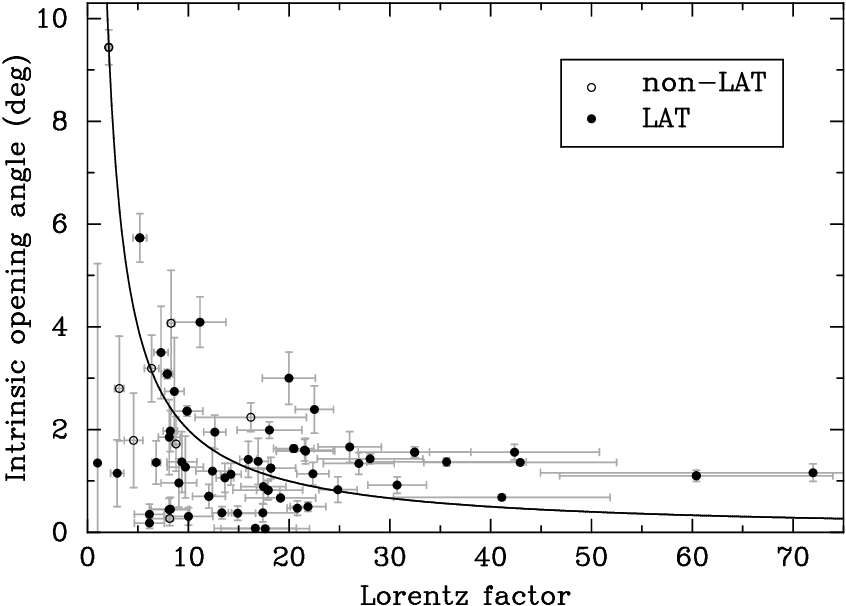}
 \caption{Intrinsic opening angle versus Lorentz factor for 65 AGN jets. The solid line shows the median
 curve fit with the assumed relation $\alpha_\text{int}=\rho/\Gamma$, where $\rho$ is a constant (here
 $\rho=0.35$ rad). Filled and open circles represent LAT-detected and non-LAT-detected sources, 
 respectively.}
 \label{f:ioa_lorentz}
\end{figure}

The intrinsic jet opening angles can be calculated as 
$\tan(\alpha_\text{int}/2)=\tan(\alpha_\text{app}/2)\sin\theta$, where $\theta$ is the viewing
angle to the jet axis. The latter, as well as the bulk Lorentz factor $\Gamma$, can be derived from 
apparent jet speed and Doppler factor using the following relations:
$$
\theta = \arctan\frac{2\beta_\text{app}}{\beta^2_\text{app}+\delta_\text{var}^2-1}\,,
\quad
\Gamma = \frac{\beta_\text{app}^2+\delta_\text{var}^2+1}{2\delta_\text{var}}\,.
$$
For $\beta_\text{app}$ and $\delta_\text{var}$ we used the fastest measured radial, non-accelerating 
apparent jet speed from the \mbox{MOJAVE} kinematic analysis \citep{MOJAVE_XIII} and the variability Doppler 
factor from the Mets\"ahovi AGN monitoring programme \citep{Hovatta09}, respectively. The corresponding 
overlap of the programmes comprises 55 sources, which are all members of the MOJAVE-1 sample. Variability 
Doppler-factors for 10 more MOJAVE-1 sources were measured within the F-GAMMA programme \citep{Liodakis17}. 
The intrinsic opening angles calculated for the 65 sources range from $0\fdg1$ to $9\fdg4$, with a 
median of $1\fdg3$, reflecting a very high degree of jet collimation. The intrinsic opening angles show 
an inverse dependence on Lorentz factor (Fig.~\ref{f:ioa_lorentz}), as predicted by hydrodynamical 
\citep{BK79} and magnetic acceleration models \citep{Komissarov07} of relativistic jets. The median value 
of the product\footnote{As we use the full opening angle, $\rho$ in this paper is not defined the same way 
as $\rho$ in \cite{Jorstad05} and \cite{Clausen-Brown13}, but they differ by a factor of 2.} 
$\rho=\alpha_\text{int}\Gamma$ is 0.35~rad, close to earlier estimates derived both from observations 
\citep{Jorstad05,Pushkarev09} and from a statistical model approach \citep{Clausen-Brown13}. The variability 
Doppler factors derived from variability can be underestimated due to a limited cadence of the observations. 
In this case, the intrinsic opening angle estimates would be smaller, while Lorentz factors would be higher 
if $\delta>(\beta_\text{app}^2+1)^{1/2}$ and smaller otherwise, implying that if the variability Doppler 
factors are essentially underestimated, the majority of points in Fig.~\ref{f:ioa_lorentz} would move 
downward and to the right. 

\begin{figure}
 \includegraphics[width=\columnwidth, angle=0]{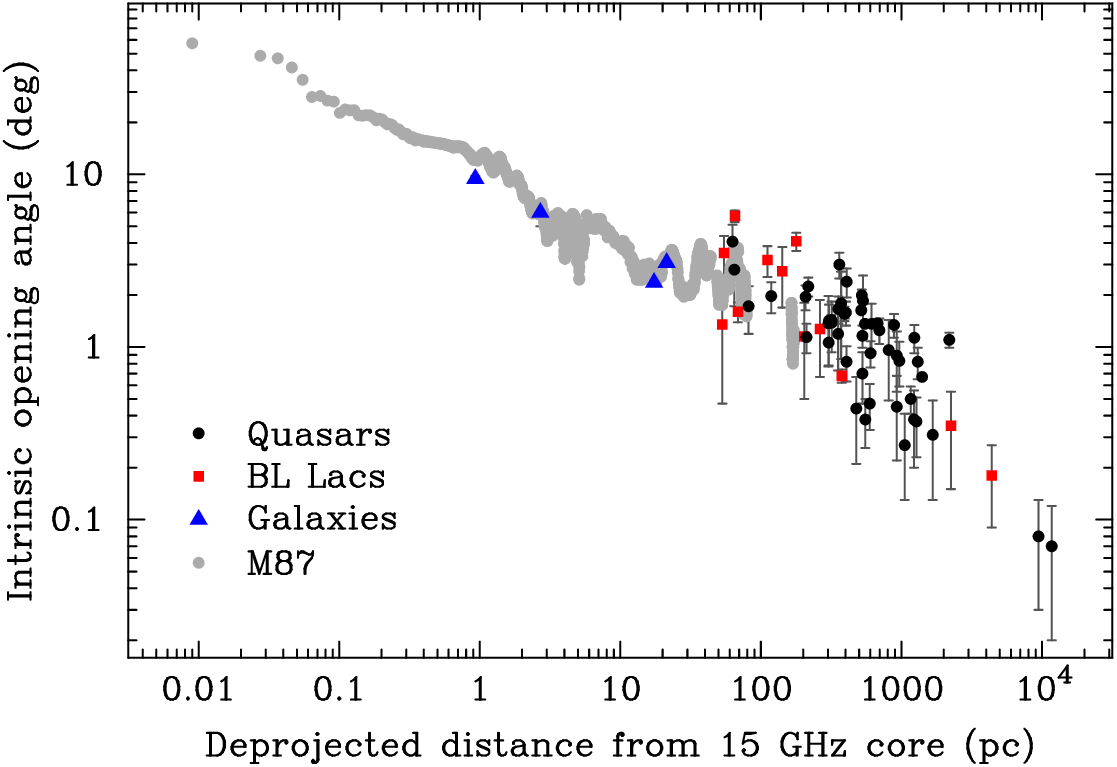}
\caption{Intrinsic opening angle versus median distance from the 15~GHz core 
         for the 65 AGN jets with viewing angle estimates (Table~\ref{t:inferred_parameters}). 
         Galaxies, BL Lacs, and quasars are shown by blue triangles, red squares, and black circles, 
         respectively. Grey points represent M87 (same data as in Fig.~\ref{f:collimation_ul_bands}, 
         right) to cover small scales.}
 \label{f:iao_r}
\end{figure}

The LAT-detected AGNs have statistically narrower jets than non-LAT-detected, with medians of
$\alpha_\text{int}^\text{LAT}=1\fdg2$ and $\alpha_\text{int}^\text{non-LAT}=2\fdg5$, respectively.
We have found no significant difference in $\alpha_\text{int}$ between quasars and BL Lacs.
The geometry and kinematics jet parameters are listed in Table~\ref{t:inferred_parameters}.

In Fig.~\ref{f:iao_r}, we plot the intrinsic jet opening angles against deprojected median distance 
from the 15~GHz core for the 65 sources supplemented by a scan along the jet of M87 (to trace small 
scales). It shows a trend from a very wide-opened outflow near the jet apex through an acceleration 
zone \citep{MOJAVE_XII,MOJAVE_XIII} to highly collimated jet regions on kiloparsec scales, with 
opening angles of the order of $1\degr$ or even narrower. Jet shapes at different scales and their 
transitions will be discussed in Kovalev et al. (in prep).

The jet viewing angle can be expressed as 
$$
\theta = \text{asin}\left(\frac{\tan(0.5\rho/\Gamma)}{\tan(0.5\,\alpha_\text{app})}\right)\approx
         \text{asin}\left(\frac{\tan(0.5\rho/\sqrt{\beta_\text{app}^2+1})}{\tan(0.5\,\alpha_\text{app})}\right)\,,
$$
where the latter approximation assumes $\theta\simeq1/\Gamma$.
We used Monte Carlo simulations to construct the viewing angle distributions separately for the LAT 
and non-LAT sub-samples, assuming flat distributions for $\rho$ in the ranges $[0.05,1.0]$ and $[0.2,0.6]$, 
respectively. The quantile functions of $\alpha_\text{app}$ and $\beta_\text{app}$ needed for the 
simulations were obtained in a closed form by applying the Generalized Lambda Distribution 
\citep[GLD:][]{GLD} technique to fit the original distributions. For $\beta_\text{app}$ we used the fastest 
radial, non-accelerating apparent jet speeds measured by the MOJAVE programme to date \citep{MOJAVE_XIII}.

\begin{figure}
 \includegraphics[width=\columnwidth, angle=0]{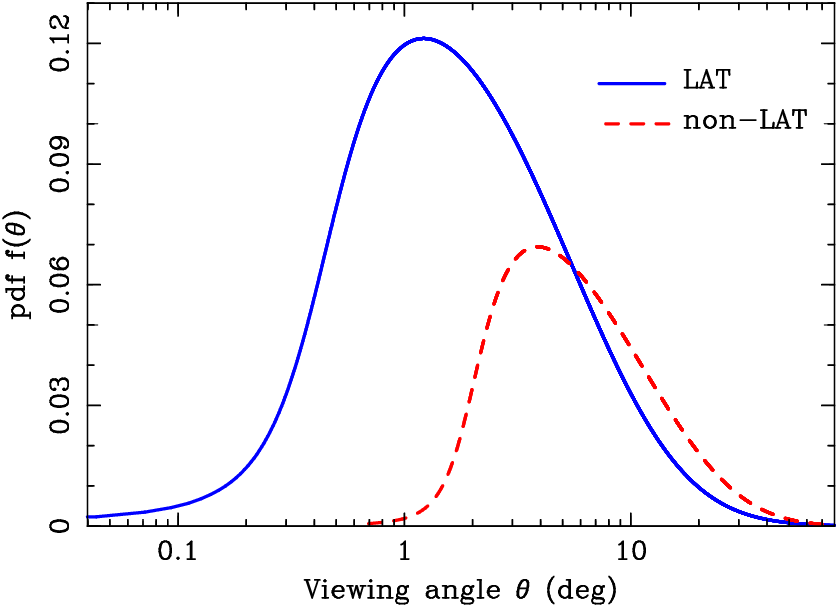}
 \caption{Probability density functions of jet viewing angle for the
          \textit{Fermi} LAT-detected (solid blue line) and non-LAT-detected (dashed red line)
          AGNs, with medians of $6\degr$ and $11\degr$, respectively.}
 \label{f:va}
\end{figure}

Applying the method of maximum likelihood and RS-parametrization of the GLD \citep{GLD_RS}, we have 
constructed probability density functions (PDFs) of the simulated viewing angle distributions for the 
LAT-detected and non-LAT-detected sources (Fig.~\ref{f:va}). The inferred PDFs have shapes close to 
log-normal and clearly show that jets of the gamma-ray bright AGNs tend to have statistically smaller 
angles to the line of sight comparing to those of gamma-ray weak AGNs, with median values $6\degr$ 
and $10\degr$, respectively. At $\theta\approx5\degr$, the probability of detecting gamma-ray weak 
source is becoming higher than that of gamma-ray bright AGN, and a ratio of the corresponding 
probabilities increases towards the right tail of the $\theta$ distributions. Statistically, the 
probability to observe a LAT-detected source within the jet cone with a typical 
$\alpha_\text{int}\approx1\degr$ is about $1.5\%$, corresponding to three sources in a sub-sample 
of 186 objects. The quasar 1520+319 (showing the widest $\alpha_\text{app}=77\degr$) might be one 
of such cases. 

We note that it is unclear whether the assumption that LAT and non-LAT detected sources have the 
same relation between the viewing angle and Lorentz factor ($\theta\simeq1/\Gamma$) introduces 
some bias in the calculation of the probability density functions of the jet viewing angle. As 
discussed by \cite{Lister15}, many AGN have not been detected by the \textit{Fermi}-LAT partly 
because of an instrumental selection effect and partly due to their lower Doppler boosting factors. 
This in some degree justifies using the same assumption for LAT-detected and non-LAT-detected 
sources. At the same time, it was shown that the approximation $\theta\simeq1/\Gamma$ could be less 
correct for jets with lower Lorentz factor and/or larger viewing angle \citep{lister_PhD}, as it is 
often the case for non-LAT-detected AGNs.

\section{Conclusions}
\label{s:summary}
We have produced total intensity stacked images at 15~GHz for 370 AGN jets having at least five epochs of
observations within the MOJAVE programme or 2~cm VLBA Survey, and constructed the corresponding ridgelines
along the outflows. Analysing projected jet width at different separations from the VLBA core for 360
sources at 15~GHz, we have found that jets of quasars and BL Lacs typicaly show a geometry close to
conical, while radio galaxies manifest streamlines closer to parabolic. The AGN jets with significant
radial accelerated motion undergo more active collimation. Jets are conical at larger, up to a few kpc,
scales probed by the 1.4~GHz VLBA observations.

By making cuts transverse to the local jet direction, we have measured widths and projected jet opening
angles on parsec scales for 362 sources (282 LAT-detected and 80 non-LAT-detected).
The apparent opening angles for $\gamma$-ray bright sources are, on average, larger than those in
$\gamma$-ray weak ones, with medians of $23\degr$ and $15\degr$, respectively. All AGNs with an apparent
jet opening angle wider than $35\degr$ are LAT-detected. We have established a highly significant correlation
between the apparent opening angle and gamma-ray luminosity, driven by Doppler beaming. We find that in many
cases, a conical jet geometry only emerges after numerous images spanning several years are stacked together.
A significant dependence is detected between the median jet width and a minimum time interval of epochs
$\Delta t$ in a stacked image if $\Delta t$ ranges from 2 to 16~yrs, with the strongest correlation at
$\Delta t\ga6$~yr, which is consistent with indications of oscillatory behaviour of jet orientation with
fitted periods ranging from 5 to 12~yrs found by \cite{MOJAVE_X}. On larger scales, probed by the single-epoch
1.4~GHz VLBA observations of a sub-sample of 122 MOJAVE sources, apparent opening angles agree well with those
at 15~GHz but on average are about 10\% wider.

The intrinsic opening angles calculated for a sample of 65 AGNs (i) range from $9\fdg4$ at scales of 
the order of 1~pc down to $0\fdg1$ at kiloparsec scales with a median of $1\fdg3$, reflecting very 
high degree of jet collimation, and (ii) show inverse dependence on Lorentz factor with a median product 
$\Gamma\alpha_\text{int}=0.35$~rad. The LAT-detected AGNs have statistically narrower jets than 
non-LAT-detected, with medians of $\alpha_\text{int,\,LAT\_Y}=1\fdg2$ and $\alpha_\text{int,LAT\_N}=2\fdg5$.
We have found no significant difference in $\alpha_\text{int}$ between quasars and BL Lacs. Performing 
Monte Carlo simulations we have constructed viewing angle distributions and showed that the gamma-ray
bright AGNs have statistically smaller angles to the line of sight comparing to those of gamma-ray weak
sources, with median values $6\degr$ and $10\degr$, respectively.

\section*{Acknowledgements}
We thank the anonymous referee for useful comments that helped to improve the manuscript.
This research has made use of data from the MOJAVE data base, which is maintained by the MOJAVE team \citep{MOJAVE_V}.
The MOJAVE project was supported by NASA-\textit{Fermi} GI grants NNX08AV67G, NNX12A087G, and NNX15AU76G.
This study was supported by the Russian Foundation for Basic Research grant 15-42-01020.
TS was funded by the Academy of Finland projects 274477 and 284495.
The National Radio Astronomy Observatory is a facility of the National Science Foundation operated under
cooperative agreement by Associated Universities, Inc.

\bibliographystyle{mnras}
\bibliography{pushkarev}

\begin{thebibliography}{}
\makeatletter
\relax
\def\mn@urlcharsother{\let\do\@makeother \do\$\do\&\do\#\do\^\do\_\do\%\do\~}
\def\mn@doi{\begingroup\mn@urlcharsother \@ifnextchar [ {\mn@doi@}
  {\mn@doi@[]}}
\def\mn@doi@[#1]#2{\def\@tempa{#1}\ifx\@tempa\@empty \href
  {http://dx.doi.org/#2} {doi:#2}\else \href {http://dx.doi.org/#2} {#1}\fi
  \endgroup}
\def\mn@eprint#1#2{\mn@eprint@#1:#2::\@nil}
\def\mn@eprint@arXiv#1{\href {http://arxiv.org/abs/#1} {{\tt arXiv:#1}}}
\def\mn@eprint@dblp#1{\href {http://dblp.uni-trier.de/rec/bibtex/#1.xml}
  {dblp:#1}}
\def\mn@eprint@#1:#2:#3:#4\@nil{\def\@tempa {#1}\def\@tempb {#2}\def\@tempc
  {#3}\ifx \@tempc \@empty \let \@tempc \@tempb \let \@tempb \@tempa \fi \ifx
  \@tempb \@empty \def\@tempb {arXiv}\fi \@ifundefined
  {mn@eprint@\@tempb}{\@tempb:\@tempc}{\expandafter \expandafter \csname
  mn@eprint@\@tempb\endcsname \expandafter{\@tempc}}}

\bibitem[\protect\citeauthoryear{{Abdo} et~al.,}{{Abdo} et~al.}{2010a}]{1FGL}
{Abdo} A.~A.,  et~al., 2010a, \mn@doi [\apjs] {10.1088/0067-0049/188/2/405},
  \href {http://adsabs.harvard.edu/abs/2010ApJS..188..405A} {188, 405}

\bibitem[\protect\citeauthoryear{{Abdo} et~al.,}{{Abdo} et~al.}{2010b}]{1LAC}
{Abdo} A.~A.,  et~al., 2010b, \mn@doi [\apj] {10.1088/0004-637X/715/1/429},
  \href {http://adsabs.harvard.edu/abs/2010ApJ...715..429A} {715, 429}

\bibitem[\protect\citeauthoryear{{Acero} et~al.,}{{Acero} et~al.}{2015}]{3FGL}
{Acero} F.,  et~al., 2015, \mn@doi [\apjs] {10.1088/0067-0049/218/2/23}, \href
  {http://adsabs.harvard.edu/abs/2015ApJS..218...23A} {218, 23}

\bibitem[\protect\citeauthoryear{{Algaba}, {Nakamura}, {Asada}  \&
  {Lee}}{{Algaba} et~al.}{2017}]{Algaba16}
{Algaba} J.~C.,  {Nakamura} M.,  {Asada} K.,   {Lee} S.~S.,  2017, \mn@doi
  [\apj] {10.3847/1538-4357/834/1/65}, \href
  {http://adsabs.harvard.edu/abs/2017ApJ...834...65A} {834, 65}

\bibitem[\protect\citeauthoryear{{Asada} \& {Nakamura}}{{Asada} \&
  {Nakamura}}{2012}]{Asada12}
{Asada} K.,  {Nakamura} M.,  2012, \mn@doi [\apjl]
  {10.1088/2041-8205/745/2/L28}, \href
  {http://adsabs.harvard.edu/abs/2012ApJ...745L..28A} {745, L28}

\bibitem[\protect\citeauthoryear{{Blandford} \& {K\"onigl}}{{Blandford} \&
  {K\"onigl}}{1979}]{BK79}
{Blandford} R.~D.,  {K\"onigl} A.,  1979, \mn@doi [\apj] {10.1086/157262},
  \href
  {http://adsabs.harvard.edu/cgi-bin/nph-bib_query?bibcode=1979ApJ...232...34B&db_key=AST}
  {232, 34}

\bibitem[\protect\citeauthoryear{{Boccardi}, {Krichbaum}, {Bach}, {Mertens},
  {Ros}, {Alef}  \& {Zensus}}{{Boccardi} et~al.}{2016}]{Boccardi15_CygnusA}
{Boccardi} B.,  {Krichbaum} T.~P.,  {Bach} U.,  {Mertens} F.,  {Ros} E.,
  {Alef} W.,   {Zensus} J.~A.,  2016, \mn@doi [\aap]
  {10.1051/0004-6361/201526985}, \href
  {http://adsabs.harvard.edu/abs/2016A%26A...585A..33B} {585, A33}

\bibitem[\protect\citeauthoryear{{B{\"o}ttcher} et~al.,}{{B{\"o}ttcher}
  et~al.}{2005}]{Bottcher05}
{B{\"o}ttcher} M.,  et~al., 2005, \mn@doi [\apj] {10.1086/432609}, \href
  {http://adsabs.harvard.edu/abs/2005ApJ...631..169B} {631, 169}

\bibitem[\protect\citeauthoryear{{Cara} \& {Lister}}{{Cara} \&
  {Lister}}{2008}]{MOJAVE_IV}
{Cara} M.,  {Lister} M.~L.,  2008, \mn@doi [\apj] {10.1086/525554}, \href
  {http://adsabs.harvard.edu/abs/2008ApJ...674..111C} {674, 111}

\bibitem[\protect\citeauthoryear{{Clausen-Brown}, {Savolainen}, {Pushkarev},
  {Kovalev}  \& {Zensus}}{{Clausen-Brown} et~al.}{2013}]{Clausen-Brown13}
{Clausen-Brown} E.,  {Savolainen} T.,  {Pushkarev} A.~B.,  {Kovalev} Y.~Y.,
  {Zensus} J.~A.,  2013, \mn@doi [\aap] {10.1051/0004-6361/201322203}, \href
  {http://adsabs.harvard.edu/abs/2013A%26A...558A.144C} {558, A144}

\bibitem[\protect\citeauthoryear{{Cohen}, {Lister}, {Homan}, {Kadler},
  {Kellermann}, {Kovalev}  \& {Vermeulen}}{{Cohen} et~al.}{2007}]{Cohen07}
{Cohen} M.~H.,  {Lister} M.~L.,  {Homan} D.~C.,  {Kadler} M.,  {Kellermann}
  K.~I.,  {Kovalev} Y.~Y.,   {Vermeulen} R.~C.,  2007, \mn@doi [\apj]
  {10.1086/511063}, \href {http://adsabs.harvard.edu/abs/2007ApJ...658..232C}
  {658, 232}

\bibitem[\protect\citeauthoryear{{Cooper}, {Lister}  \& {Kochanczyk}}{{Cooper}
  et~al.}{2007}]{MOJAVE_III}
{Cooper} N.~J.,  {Lister} M.~L.,   {Kochanczyk} M.~D.,  2007, \mn@doi [\apjs]
  {10.1086/518654}, \href {http://adsabs.harvard.edu/abs/2007ApJS..171..376C}
  {171, 376}

\bibitem[\protect\citeauthoryear{{Giroletti}, {Giovannini}, {Cotton}, {Taylor},
  {P{\'e}rez-Torres}, {Chiaberge}  \& {Edwards}}{{Giroletti}
  et~al.}{2008}]{Giroletti08_Mkn501}
{Giroletti} M.,  {Giovannini} G.,  {Cotton} W.~D.,  {Taylor} G.~B.,
  {P{\'e}rez-Torres} M.~A.,  {Chiaberge} M.,   {Edwards} P.~G.,  2008, \mn@doi
  [\aap] {10.1051/0004-6361:200809784}, \href
  {http://adsabs.harvard.edu/abs/2008A%26A...488..905G} {488, 905}

\bibitem[\protect\citeauthoryear{{Hada}, {Doi}, {Kino}, {Nagai}, {Hagiwara}  \&
  {Kawaguchi}}{{Hada} et~al.}{2011}]{Hada11_M87}
{Hada} K.,  {Doi} A.,  {Kino} M.,  {Nagai} H.,  {Hagiwara} Y.,   {Kawaguchi}
  N.,  2011, \mn@doi [\nat] {10.1038/nature10387}, \href
  {http://adsabs.harvard.edu/abs/2011Natur.477..185H} {477, 185}

\bibitem[\protect\citeauthoryear{{Hada} et~al.,}{{Hada}
  et~al.}{2013}]{Hada13_M87}
{Hada} K.,  et~al., 2013, \mn@doi [\apj] {10.1088/0004-637X/775/1/70}, \href
  {http://adsabs.harvard.edu/abs/2013ApJ...775...70H} {775, 70}

\bibitem[\protect\citeauthoryear{{Hada} et~al.,}{{Hada}
  et~al.}{2016}]{Hada16_M87}
{Hada} K.,  et~al., 2016, \mn@doi [\apj] {10.3847/0004-637X/817/2/131}, \href
  {http://adsabs.harvard.edu/abs/2016ApJ...817..131H} {817, 131}

\bibitem[\protect\citeauthoryear{{Homan} \& {Lister}}{{Homan} \&
  {Lister}}{2006}]{MOJAVE_II}
{Homan} D.~C.,  {Lister} M.~L.,  2006, \mn@doi [\aj] {10.1086/500256}, \href
  {http://adsabs.harvard.edu/abs/2006AJ....131.1262H} {131, 1262}

\bibitem[\protect\citeauthoryear{{Homan}, {Kadler}, {Kellermann}, {Kovalev},
  {Lister}, {Ros}, {Savolainen}  \& {Zensus}}{{Homan}
  et~al.}{2009}]{MOJAVE_VII}
{Homan} D.~C.,  {Kadler} M.,  {Kellermann} K.~I.,  {Kovalev} Y.~Y.,  {Lister}
  M.~L.,  {Ros} E.,  {Savolainen} T.,   {Zensus} J.~A.,  2009, \mn@doi [\apj]
  {10.1088/0004-637X/706/2/1253}, \href
  {http://adsabs.harvard.edu/abs/2009ApJ...706.1253H} {706, 1253}

\bibitem[\protect\citeauthoryear{{Homan}, {Lister}, {Kovalev}, {Pushkarev},
  {Savolainen}, {Kellermann}, {Richards}  \& {Ros}}{{Homan}
  et~al.}{2015}]{MOJAVE_XII}
{Homan} D.~C.,  {Lister} M.~L.,  {Kovalev} Y.~Y.,  {Pushkarev} A.~B.,
  {Savolainen} T.,  {Kellermann} K.~I.,  {Richards} J.~L.,   {Ros} E.,  2015,
  \mn@doi [\apj] {10.1088/0004-637X/798/2/134}, \href
  {http://adsabs.harvard.edu/abs/2015ApJ...798..134H} {798, 134}

\bibitem[\protect\citeauthoryear{{Hovatta}, {Valtaoja}, {Tornikoski}  \&
  {L{\"a}hteenm{\"a}ki}}{{Hovatta} et~al.}{2009}]{Hovatta09}
{Hovatta} T.,  {Valtaoja} E.,  {Tornikoski} M.,   {L{\"a}hteenm{\"a}ki} A.,
  2009, \mn@doi [\aap] {10.1051/0004-6361/200811150e}, \href
  {http://adsabs.harvard.edu/abs/2009A%26A...498..723H} {498, 723}

\bibitem[\protect\citeauthoryear{{Hovatta}, {Lister}, {Aller}, {Aller},
  {Homan}, {Kovalev}, {Pushkarev}  \& {Savolainen}}{{Hovatta}
  et~al.}{2012}]{MOJAVE_VIII}
{Hovatta} T.,  {Lister} M.~L.,  {Aller} M.~F.,  {Aller} H.~D.,  {Homan} D.~C.,
  {Kovalev} Y.~Y.,  {Pushkarev} A.~B.,   {Savolainen} T.,  2012, \mn@doi [\aj]
  {10.1088/0004-6256/144/4/105}, \href
  {http://adsabs.harvard.edu/abs/2012AJ....144..105H} {144, 105}

\bibitem[\protect\citeauthoryear{{Hovatta} et~al.,}{{Hovatta}
  et~al.}{2014}]{MOJAVE_XI}
{Hovatta} T.,  et~al., 2014, \mn@doi [\aj] {10.1088/0004-6256/147/6/143}, \href
  {http://adsabs.harvard.edu/abs/2014AJ....147..143H} {147, 143}

\bibitem[\protect\citeauthoryear{{Jones}, {Saunders}, {Read}  \&
  {Colless}}{{Jones} et~al.}{2005}]{2005PASA...22..277J}
{Jones} D.~H.,  {Saunders} W.,  {Read} M.,   {Colless} M.,  2005, \mn@doi
  [\pasa] {10.1071/AS05018}, \href
  {http://adsabs.harvard.edu/abs/2005PASA...22..277J} {22, 277}

\bibitem[\protect\citeauthoryear{{Jorstad} et~al.,}{{Jorstad}
  et~al.}{2005}]{Jorstad05}
{Jorstad} S.~G.,  et~al., 2005, \mn@doi [\aj] {10.1086/444593}, \href
  {http://adsabs.harvard.edu/cgi-bin/nph-bib_query?bibcode=2005AJ....130.1418J&db_key=AST}
  {130, 1418}

\bibitem[\protect\citeauthoryear{{Junor}, {Biretta}  \& {Livio}}{{Junor}
  et~al.}{1999}]{Junor99_M87}
{Junor} W.,  {Biretta} J.~A.,   {Livio} M.,  1999, \mn@doi [\nat]
  {10.1038/44780}, \href {http://adsabs.harvard.edu/abs/1999Natur.401..891J}
  {401, 891}

\bibitem[\protect\citeauthoryear{{Kadler}, {Ros}, {Lobanov}, {Falcke}  \&
  {Zensus}}{{Kadler} et~al.}{2004}]{Kadler04}
{Kadler} M.,  {Ros} E.,  {Lobanov} A.~P.,  {Falcke} H.,   {Zensus} J.~A.,
  2004, \mn@doi [\aap] {10.1051/0004-6361:20041051}, \href
  {http://adsabs.harvard.edu/abs/2004A%26A...426..481K} {426, 481}

\bibitem[\protect\citeauthoryear{{Kellermann} \& {Pauliny-Toth}}{{Kellermann}
  \& {Pauliny-Toth}}{1969}]{Kellermann69}
{Kellermann} K.~I.,  {Pauliny-Toth} I.~I.~K.,  1969, \mn@doi [\apjl]
  {10.1086/180305}, \href {http://adsabs.harvard.edu/abs/1969ApJ...155L..71K}
  {155, L71}

\bibitem[\protect\citeauthoryear{{Kellermann}, {Vermeulen}, {Zensus}  \&
  {Cohen}}{{Kellermann} et~al.}{1998}]{2cmPaperI}
{Kellermann} K.~I.,  {Vermeulen} R.~C.,  {Zensus} J.~A.,   {Cohen} M.~H.,
  1998, \mn@doi [AJ] {10.1086/300308}, \href
  {http://adsabs.harvard.edu/cgi-bin/nph-bib_query?bibcode=1998AJ....115.1295K&db_key=AST}
  {115, 1295}

\bibitem[\protect\citeauthoryear{{King} \& {MacGillivray}}{{King} \&
  {MacGillivray}}{1999}]{GLD}
{King} R.,  {MacGillivray} H.,  1999, Australian and New Zealand Journal of
  Statistics, 41

\bibitem[\protect\citeauthoryear{{Komatsu} et~al.,}{{Komatsu}
  et~al.}{2009}]{Komatsu09}
{Komatsu} E.,  et~al., 2009, \mn@doi [\apjs] {10.1088/0067-0049/180/2/330},
  \href {http://adsabs.harvard.edu/abs/2009ApJS..180..330K} {180, 330}

\bibitem[\protect\citeauthoryear{{Komissarov}, {Barkov}, {Vlahakis}  \&
  {K{\"o}nigl}}{{Komissarov} et~al.}{2007}]{Komissarov07}
{Komissarov} S.~S.,  {Barkov} M.~V.,  {Vlahakis} N.,   {K{\"o}nigl} A.,  2007,
  \mn@doi [\mnras] {10.1111/j.1365-2966.2007.12050.x}, \href
  {http://adsabs.harvard.edu/abs/2007MNRAS.380...51K} {380, 51}

\bibitem[\protect\citeauthoryear{{Kovalev}, {Lobanov}, {Pushkarev}  \&
  {Zensus}}{{Kovalev} et~al.}{2008}]{Kovalev_08_cs}
{Kovalev} Y.~Y.,  {Lobanov} A.~P.,  {Pushkarev} A.~B.,   {Zensus} J.~A.,  2008,
  \mn@doi [\aap] {10.1051/0004-6361:20078679}, \href
  {http://adsabs.harvard.edu/abs/2008A%26A...483..759K} {483, 759}

\bibitem[\protect\citeauthoryear{{Kovalev}, {Petrov}  \& {Plavin}}{{Kovalev}
  et~al.}{2017}]{Kovalev17}
{Kovalev} Y.~Y.,  {Petrov} L.,   {Plavin} A.~V.,  2017, \mn@doi [\aap]
  {10.1051/0004-6361/201630031}, \href
  {http://adsabs.harvard.edu/abs/2017A%26A...598L...1K} {598, L1}

\bibitem[\protect\citeauthoryear{{Kravchenko}, {Kovalev}, {Hovatta}  \&
  {Ramakrishnan}}{{Kravchenko} et~al.}{2016}]{Kravchenko16}
{Kravchenko} E.~V.,  {Kovalev} Y.~Y.,  {Hovatta} T.,   {Ramakrishnan} V.,
  2016, \mn@doi [\mnras] {10.1093/mnras/stw1776}, \href
  {http://adsabs.harvard.edu/abs/2016MNRAS.462.2747K} {462, 2747}

\bibitem[\protect\citeauthoryear{{Kutkin} et~al.,}{{Kutkin}
  et~al.}{2014}]{Kutkin14}
{Kutkin} A.~M.,  et~al., 2014, \mn@doi [\mnras] {10.1093/mnras/stt2133}, \href
  {http://adsabs.harvard.edu/abs/2014MNRAS.437.3396K} {437, 3396}

\bibitem[\protect\citeauthoryear{{Lawrence}, {Pearson}, {Readhead}  \&
  {Unwin}}{{Lawrence} et~al.}{1986}]{1986AJ.....91..494L}
{Lawrence} C.~R.,  {Pearson} T.~J.,  {Readhead} A.~C.~S.,   {Unwin} S.~C.,
  1986, \mn@doi [\aj] {10.1086/114027}, \href
  {http://adsabs.harvard.edu/abs/1986AJ.....91..494L} {91, 494}

\bibitem[\protect\citeauthoryear{{Liodakis} et~al.,}{{Liodakis}
  et~al.}{2017}]{Liodakis17}
{Liodakis} I.,  et~al., 2017, \mn@doi [\mnras] {10.1093/mnras/stx002}, \href
  {http://adsabs.harvard.edu/abs/2017MNRAS.466.4625L} {466, 4625}

\bibitem[\protect\citeauthoryear{{Lister}}{{Lister}}{1999}]{lister_PhD}
{Lister} M.~L.,  1999, PhD thesis, BOSTON UNIVERSITY

\bibitem[\protect\citeauthoryear{{Lister} \& {Homan}}{{Lister} \&
  {Homan}}{2005}]{MOJAVE_I}
{Lister} M.~L.,  {Homan} D.~C.,  2005, \mn@doi [\aj] {10.1086/432969}, \href
  {http://adsabs.harvard.edu/abs/2005AJ....130.1389L} {130, 1389}

\bibitem[\protect\citeauthoryear{{Lister} et~al.,}{{Lister}
  et~al.}{2009a}]{MOJAVE_V}
{Lister} M.~L.,  et~al., 2009a, \mn@doi [\aj] {10.1088/0004-6256/137/3/3718},
  \href {http://adsabs.harvard.edu/abs/2009AJ....137.3718L} {137, 3718}

\bibitem[\protect\citeauthoryear{{Lister} et~al.,}{{Lister}
  et~al.}{2009b}]{MOJAVE_VI}
{Lister} M.~L.,  et~al., 2009b, \mn@doi [\aj] {10.1088/0004-6256/138/6/1874},
  \href {http://adsabs.harvard.edu/abs/2009AJ....138.1874L} {138, 1874}

\bibitem[\protect\citeauthoryear{{Lister} et~al.,}{{Lister}
  et~al.}{2011}]{Lister11}
{Lister} M.~L.,  et~al., 2011, \mn@doi [\apj] {10.1088/0004-637X/742/1/27},
  \href {http://adsabs.harvard.edu/abs/2011ApJ...742...27L} {742, 27}

\bibitem[\protect\citeauthoryear{{Lister} et~al.,}{{Lister}
  et~al.}{2013}]{MOJAVE_X}
{Lister} M.~L.,  et~al., 2013, \mn@doi [\aj] {10.1088/0004-6256/146/5/120},
  \href {http://adsabs.harvard.edu/abs/2013AJ....146..120L} {146, 120}

\bibitem[\protect\citeauthoryear{{Lister}, {Aller}, {Aller}, {Hovatta},
  {Max-Moerbeck}, {Readhead}, {Richards}  \& {Ros}}{{Lister}
  et~al.}{2015}]{Lister15}
{Lister} M.~L.,  {Aller} M.~F.,  {Aller} H.~D.,  {Hovatta} T.,  {Max-Moerbeck}
  W.,  {Readhead} A.~C.~S.,  {Richards} J.~L.,   {Ros} E.,  2015, \mn@doi
  [\apjl] {10.1088/2041-8205/810/1/L9}, \href
  {http://adsabs.harvard.edu/abs/2015ApJ...810L...9L} {810, L9}

\bibitem[\protect\citeauthoryear{{Lister} et~al.,}{{Lister}
  et~al.}{2016}]{MOJAVE_XIII}
{Lister} M.~L.,  et~al., 2016, \mn@doi [\aj] {10.3847/0004-6256/152/1/12},
  \href {http://adsabs.harvard.edu/abs/2016AJ....152...12L} {152, 12}

\bibitem[\protect\citeauthoryear{{Lobanov}}{{Lobanov}}{1998}]{Lobanov_98}
{Lobanov} A.~P.,  1998, \aap, \href
  {http://adsabs.harvard.edu/cgi-bin/nph-bib_query?bibcode=1998A%26A...330...79L&db_key=AST}
  {330, 79}

\bibitem[\protect\citeauthoryear{{Nakamura} \& {Asada}}{{Nakamura} \&
  {Asada}}{2013}]{Nakamura13_M87}
{Nakamura} M.,  {Asada} K.,  2013, \mn@doi [\apj]
  {10.1088/0004-637X/775/2/118}, \href
  {http://adsabs.harvard.edu/abs/2013ApJ...775..118N} {775, 118}

\bibitem[\protect\citeauthoryear{{Nolan} et~al.,}{{Nolan} et~al.}{2012}]{2FGL}
{Nolan} P.~L.,  et~al., 2012, \mn@doi [\apjs] {10.1088/0067-0049/199/2/31},
  \href {http://adsabs.harvard.edu/abs/2012ApJS..199...31N} {199, 31}

\bibitem[\protect\citeauthoryear{{Ojha} et~al.,}{{Ojha} et~al.}{2010}]{Ojha10}
{Ojha} R.,  et~al., 2010, \mn@doi [\aap] {10.1051/0004-6361/200912724}, \href
  {http://adsabs.harvard.edu/abs/2010A%26A...519A..45O} {519, A45}

\bibitem[\protect\citeauthoryear{{Pushkarev} \& {Kovalev}}{{Pushkarev} \&
  {Kovalev}}{2012}]{RDV_paper}
{Pushkarev} A.~B.,  {Kovalev} Y.~Y.,  2012, \mn@doi [\aap]
  {10.1051/0004-6361/201219352}, \href
  {http://adsabs.harvard.edu/abs/2012A%26A...544A..34P} {544, A34}

\bibitem[\protect\citeauthoryear{{Pushkarev} \& {Kovalev}}{{Pushkarev} \&
  {Kovalev}}{2015}]{Pushkarev15}
{Pushkarev} A.~B.,  {Kovalev} Y.~Y.,  2015, \mn@doi [\mnras]
  {10.1093/mnras/stv1539}, \href
  {http://adsabs.harvard.edu/abs/2015MNRAS.452.4274P} {452, 4274}

\bibitem[\protect\citeauthoryear{{Pushkarev}, {Kovalev}, {Lister}  \&
  {Savolainen}}{{Pushkarev} et~al.}{2009}]{Pushkarev09}
{Pushkarev} A.~B.,  {Kovalev} Y.~Y.,  {Lister} M.~L.,   {Savolainen} T.,  2009,
  \mn@doi [\aap] {10.1051/0004-6361/200913422}, \href
  {http://adsabs.harvard.edu/abs/2009A%26A...507L..33P} {507, L33}

\bibitem[\protect\citeauthoryear{{Pushkarev}, {Hovatta}, {Kovalev}, {Lister},
  {Lobanov}, {Savolainen}  \& {Zensus}}{{Pushkarev} et~al.}{2012}]{MOJAVE_IX}
{Pushkarev} A.~B.,  {Hovatta} T.,  {Kovalev} Y.~Y.,  {Lister} M.~L.,  {Lobanov}
  A.~P.,  {Savolainen} T.,   {Zensus} J.~A.,  2012, \mn@doi [\aap]
  {10.1051/0004-6361/201219173}, \href
  {http://adsabs.harvard.edu/abs/2012A%26A...545A.113P} {545, A113}

\bibitem[\protect\citeauthoryear{{Pushkarev} et~al.,}{{Pushkarev}
  et~al.}{2013}]{Pushkarev13}
{Pushkarev} A.~B.,  et~al., 2013, \mn@doi [\aap] {10.1051/0004-6361/201321484},
  \href {http://adsabs.harvard.edu/abs/2013A%26A...555A..80P} {555, A80}

\bibitem[\protect\citeauthoryear{{Ramberg} \& {Schmeiser}}{{Ramberg} \&
  {Schmeiser}}{1974}]{GLD_RS}
{Ramberg} J.,  {Schmeiser} B.,  1974, Communications of the Association for
  Computing Machinery, 17, 78

\bibitem[\protect\citeauthoryear{{Rau} et~al.,}{{Rau}
  et~al.}{2012}]{2012AA...538A..26R}
{Rau} A.,  et~al., 2012, \mn@doi [\aap] {10.1051/0004-6361/201118159}, \href
  {http://adsabs.harvard.edu/abs/2012A%26A...538A..26R} {538, A26}

\bibitem[\protect\citeauthoryear{{Sargent}}{{Sargent}}{1970}]{1970ApJ...160..405S}
{Sargent} W.~L.~W.,  1970, \mn@doi [\apj] {10.1086/150443}, \href
  {http://adsabs.harvard.edu/abs/1970ApJ...160..405S} {160, 405}

\bibitem[\protect\citeauthoryear{{Savolainen}, {Wiik}, {Valtaoja}, {Kadler},
  {Ros}, {Tornikoski}, {Aller}  \& {Aller}}{{Savolainen}
  et~al.}{2006}]{Savolainen06}
{Savolainen} T.,  {Wiik} K.,  {Valtaoja} E.,  {Kadler} M.,  {Ros} E.,
  {Tornikoski} M.,  {Aller} M.~F.,   {Aller} H.~D.,  2006, \mn@doi [\apj]
  {10.1086/505259}, \href {http://adsabs.harvard.edu/abs/2006ApJ...647..172S}
  {647, 172}

\bibitem[\protect\citeauthoryear{{Shaw} et~al.,}{{Shaw}
  et~al.}{2012}]{2012ApJ...748...49S}
{Shaw} M.~S.,  et~al., 2012, \mn@doi [\apj] {10.1088/0004-637X/748/1/49}, \href
  {http://adsabs.harvard.edu/abs/2012ApJ...748...49S} {748, 49}

\bibitem[\protect\citeauthoryear{{Shaw} et~al.,}{{Shaw}
  et~al.}{2013}]{2013ApJ...764..135S}
{Shaw} M.~S.,  et~al., 2013, \mn@doi [\apj] {10.1088/0004-637X/764/2/135},
  \href {http://adsabs.harvard.edu/abs/2013ApJ...764..135S} {764, 135}

\bibitem[\protect\citeauthoryear{{Shepherd}}{{Shepherd}}{1997}]{difmap}
{Shepherd} M.~C.,  1997, in {Hunt} G.,  {Payne} H.~E.,  eds,  Astronomical
  Society of the Pacific Conference Series Vol. 125, Astronomical Data Analysis
  Software and Systems VI. San Francisco: ASP, p.~77

\bibitem[\protect\citeauthoryear{{Sokolovsky}, {Kovalev}, {Pushkarev}  \&
  {Lobanov}}{{Sokolovsky} et~al.}{2011}]{Sokolovsky_11}
{Sokolovsky} K.~V.,  {Kovalev} Y.~Y.,  {Pushkarev} A.~B.,   {Lobanov} A.~P.,
  2011, \mn@doi [\aap] {10.1051/0004-6361/201016072}, \href
  {http://adsabs.harvard.edu/abs/2011A%26A...532A..38S} {532, A38}

\bibitem[\protect\citeauthoryear{{Stickel} \& {Kuehr}}{{Stickel} \&
  {Kuehr}}{1994}]{1994AAS..103..349S}
{Stickel} M.,  {Kuehr} H.,  1994, \aaps, \href
  {http://adsabs.harvard.edu/abs/1994A%26AS..103..349S} {103}

\bibitem[\protect\citeauthoryear{{Thompson}, {Djorgovski}, {Vigotti}  \&
  {Grueff}}{{Thompson} et~al.}{1992}]{1992ApJS...81....1T}
{Thompson} D.~J.,  {Djorgovski} S.,  {Vigotti} M.,   {Grueff} G.,  1992,
  \mn@doi [\apjs] {10.1086/191683}, \href
  {http://adsabs.harvard.edu/abs/1992ApJS...81....1T} {81, 1}

\bibitem[\protect\citeauthoryear{{Tseng}, {Asada}, {Nakamura}, {Pu}, {Algaba}
  \& {Lo}}{{Tseng} et~al.}{2016}]{Tseng16}
{Tseng} C.-Y.,  {Asada} K.,  {Nakamura} M.,  {Pu} H.-Y.,  {Algaba} J.-C.,
  {Lo} W.-P.,  2016, \mn@doi [\apj] {10.3847/1538-4357/833/2/288}, \href
  {http://adsabs.harvard.edu/abs/2016ApJ...833..288T} {833, 288}

\bibitem[\protect\citeauthoryear{{Vermeulen} \& {Cohen}}{{Vermeulen} \&
  {Cohen}}{1994}]{Vermeulen94}
{Vermeulen} R.~C.,  {Cohen} M.~H.,  1994, \mn@doi [\apj] {10.1086/174424},
  \href {http://adsabs.harvard.edu/abs/1994ApJ...430..467V} {430, 467}

\bibitem[\protect\citeauthoryear{{Zensus}, {Ros}, {Kellermann}, {Cohen},
  {Vermeulen}  \& {Kadler}}{{Zensus} et~al.}{2002}]{2002AJ....124..662Z}
{Zensus} J.~A.,  {Ros} E.,  {Kellermann} K.~I.,  {Cohen} M.~H.,  {Vermeulen}
  R.~C.,   {Kadler} M.,  2002, \mn@doi [\aj] {10.1086/341585}, \href
  {http://adsabs.harvard.edu/abs/2002AJ....124..662Z} {124, 662}

\makeatother
\end{thebibliography}

% Don't change these lines
\bsp    % typesetting comment
\label{lastpage}

\begin{landscape}
\begin{table}
%\begin{minipage}{220mm}
\caption{Collimation and kinematic source parameters. Columns are as follows:
(1)  B1950 name;
(2)  apparent opening angle at 15~GHz;
(3)  $k$-index at 15~GHz;
(4)  angular jet length along ridgeline at 15~GHz;
(5)  linear projected jet length at 15~GHz;
(6)  maximum apparent speed at 15~GHz from Lister et al. (2016);
(7)  variability Doppler factor from Hovatta et al. (2009) and Liodakis et al. (2017);
(8)  Lorentz factor;
(9)  viewing angle;
(10) intrinsic opening angle;
(11) $k$-index at 2~GHz;
(12) angular jet length along ridgeline at 2~GHz;
(13) linear projected jet length at 2~GHz; and
(14) linear deprojected jet length at 2~GHz.
This table is available in its entirety in the online journal.
A portion is shown here for guidance regarding its form and content.}
\label{t:inferred_parameters}
\begin{tabular}{c r c r r c c c c c c c c c}
\hline
\noalign{\smallskip}
 B1950   & $\alpha_\text{app}^{15}$ & $k_{15}$ & $r_{15}$ &$l_{15}$ & $\beta_\text{app}$ & $\delta$ & $\Gamma$ & $\theta$ & $\alpha_\text{int}$ & $k_2$ & $r_2$ & $l_2$ & $L_2$ \\
           & (deg) &       & (mas) &  (pc)  &   (c)  &       &       & (deg) & (deg) &       &  (mas) &  (kpc) & (kpc) \\
   (1)     &  (2)  &  (3)  &  (4)  &   (5)  &   (6)  &  (7)  &  (8)  &  (9)  &  (10) &  (11) &   (12) &   (13) &  (14) \\
\hline\noalign{\smallskip}
0003$-$066 & $16.5\pm1.8$ & $0.44\pm0.03$ &  8.8 &   42.9 &  $6.15\pm0.54$ & $5.1\pm1.3$ &  $6.4\pm0.7$ & $11.1\pm2.2$ & $3.2\pm0.7$ & $0.53\pm0.03$ &   21.3 &   0.10 &   0.5 \\
0003$+$380 & $27.3\pm4.3$ & $1.76\pm0.08$ &  4.9 &   17.8 &         \ldots &      \ldots &       \ldots &       \ldots &      \ldots &        \ldots & \ldots & \ldots &\ldots \\
0006$+$061 & $13.3\pm2.9$ & $0.69\pm0.03$ & 22.4 & \ldots &         \ldots &      \ldots &       \ldots &       \ldots &      \ldots &        \ldots & \ldots & \ldots &\ldots \\
0007$+$106 & $23.2\pm3.7$ & $1.91\pm0.06$ &  1.5 &    2.5 &         \ldots &      \ldots &       \ldots &       \ldots &      \ldots &        \ldots & \ldots & \ldots &\ldots \\
0010$+$405 &  $9.3\pm0.6$ & $0.85\pm0.07$ &  9.1 &   35.9 &         \ldots &      \ldots &       \ldots &       \ldots &      \ldots &        \ldots & \ldots & \ldots &\ldots \\
0011$+$189 & $15.6\pm1.1$ & $0.67\pm0.03$ &  1.9 &   11.3 &         \ldots &      \ldots &       \ldots &       \ldots &      \ldots &        \ldots & \ldots & \ldots &\ldots \\
0015$-$054 & $13.2\pm1.1$ & $0.52\pm0.03$ &  2.6 &    9.3 &         \ldots &      \ldots &       \ldots &       \ldots &      \ldots &        \ldots & \ldots & \ldots &\ldots \\
0016$+$731 & $31.0\pm2.5$ & $1.28\pm0.02$ &  1.9 &   16.2 &  $8.22\pm0.31$ & $7.8\pm1.9$ &  $8.3\pm0.3$ &  $7.4\pm1.7$ & $4.1\pm1.0$ & $0.89\pm0.02$ &   73.7 &   0.63 &   4.9 \\
0019$+$058 & $43.6\pm3.8$ & $0.69\pm0.13$ &  3.5 & \ldots &         \ldots &      \ldots &       \ldots &       \ldots &      \ldots &        \ldots & \ldots & \ldots &\ldots \\
\hline
\end{tabular}
%\end{minipage}
\end{table}
\end{landscape}

\end{document}